\documentclass[12pt]{iopart}
\usepackage{graphicx}
\usepackage{subcaption}
\usepackage{hyperref}
\usepackage{stackengine}
\usepackage[OT1,T2A]{fontenc}
\eqnobysec
\newcommand{\ba}{\beta}
\begin{document}
\title{Exact solutions for charged spheres and their stability. II. Anisotropic Fluids.}

\author{ K. Dev}

\address{Department of Physics and Astronomy, Dickinson College, Carlisle PA }
\ead{devk@dickinson.edu}

\begin{abstract}
We study exact solutions of the Einstein-Maxwell equations for the interior gravitational field of static spherically symmetric charged compact spheres. The spheres consist of an anisotropic fluid  with a charge distribution that gives rise to a static radial electric field. The density  of the fluid has the form $\rho (r) = \rho_o + \alpha r^2$ (here  $\rho_o$ and $\alpha$ are constants) and the total charge $q(r)$ within a sphere of radius $r$ has the form $q = \beta r^3$ (with $\beta$ a constant). We evaluate the critical values of $M/R$ for these spheres as a function of $Q/R$ and compare these values with those given by the Andr\'{e}asson formula. 
\end{abstract}


\section{Introduction}
In a recent paper \cite{Dev1}, we studied exact solutions of the Einstein-Maxwell equations for the interior gravitational field of static spherically symmetric charged perfect fluid compact objects. Here, we extend our study to include charged spheres with anisotropic pressure. The term anisotropic pressure is used to describe physical situations in which the pressures associated with different directions differ from each other.  
Anisotropic spheres are compact distributions of matter  in which the radial pressure is not equal to the tangential pressure,  $p_r \neq p_t$.

The pioneering study of the effects of anisotropic pressure on the properties of compact spheres was done by Bowers and Liang \cite{Bowers}. The Tolman-Oppenheimer-Volkoff (TOV) equation for an anisotropic sphere with constant density, $\rho = \rho_0 = const$ can be written as 
\begin{equation}
\label{eon2}
\frac {dp_r}{dr}=  \frac{2}{r}(p_t - p_r)  - \frac{4 \pi r\left(p_r^2 + \frac{4}{3} p_r \rho_o + \frac{1}{3} \rho_0^2 \right)}{\left( 1 - \frac{8}{3} \pi \rho_0 r^2\right)}
\end{equation}
In the case of  isotropic spheres where $p_r = p_t$ this equation can be integrated. Bowers and Liang proposed for anisotropic spheres with $\rho = \rho_0 = const$ the following ansatz: 
\begin{equation}
 \frac{2}{r}(p_t - p_r)  = C \frac{r^2\left(p_r^2 + \frac{4}{3} p_r \rho_o + \frac{1}{3} \rho_0^2 \right)}{\left( 1 - \frac{8}{3} \pi \rho_0 r^2\right)}
\end{equation}
They were then able to integrate (\ref{eon2}) and obtain an expression for $p_r$. They studied the resulting equation for $p_r$ and showed that the maximum  value of $M/R$ (where $M$ is total mass and $R$ is the radius of the sphere) for anisotropic spheres can be greater than $4/9$ before gravitational collapse occurs. Buchdahl \cite{Buchdahl}, had earlier shown that gravitational collapse will occur if $M/R$ for a compact object with spherical symmetry exceeds $4/9$. However, it must be noted that Buchdahl studied perfect fluid spheres. 

Since Bowers and Liang's  paper many  relativists have contributed to the study and development of the physics of anisotropic spheres. Bayin \cite{Bayin}, considered   generalizations of the perfect fluid solution with $ p = \alpha \rho$  to the anisotropic case. Herrera and his co-workers   \cite{Herrera1}-\cite{Herrera5},              have studied conformally flat anisotropic spheres, dissipative anisotropic fluids, anisotropic double polytrope fluids, cracking of anisotropic polytropes,  anisotropic geodesic fluids and shear-free anisotropic fluids.  Boonserm et al. \cite{Visser}, have shown that an anisotropic fluid can be modeled as  a classical (charged) isotropic perfect fluid,
a classical electromagnetic field and a classical (minimally coupled) scalar field.  There have been at least two articles that claim to have   developed a formalism for generating all anisotropic solutions, \cite{Akbar} and  \cite{Herrera6}.


In this paper we are studying interior solutions  for  charged anisotropic spheres. These solutions must match the  Reissner-Nordstr\"{o}m  metric (Reissner \cite{Reissner}, Weyl  \cite{Weyl} and Nordstr\"{o}m   \cite{Nordstrom})  at the surface of the sphere. The  Reissner-Nordstr\"{o}m  metric is the solution of the Einstein-Maxwell equations that represents the exterior gravitational field of a spherically symmetric charged body.  In fact,  it is the unique asymptotically flat vacuum solution around any
charged spherically symmetric object, irrespective of how that body may be composed
or how it may evolve in time  (Carter \cite{Carter}, Ruback  \cite{Ruback} and Chru\'{s}ciel  \cite{Chrusciel}).

The internal structure of a charged anisotropic sphere is established by solving the coupled Einstein-Maxwell equations. A solution of these equations  establishes  the gravitational field inside the sphere as a function of the distribution of matter and energy within the sphere. The Einstein-Maxwell equations for a charged static anisotropic spherically distribution of matter  reduce to a set  of four independent non-linear second order differential equations that connect  the metric coefficients $g_{tt}$ and $g_{rr}$ with the physical quantities that represent the matter content of the sphere.  The matter content of charged anisotropic  fluid is given  by functions that describes the inertial density $\rho_{fl}$, the radial pressure $p_r$, the tangential pressure $p_t$ and the electromagnetic energy density $\rho_{em}$ or the electromagnetic charge distribution $q(r)$. The complete system of equations to be solved consists of four linearly independent equations with six unknowns.

An important question that arises in the study of the structure of  spherically symmetric compact objects  is the following: what is the maximum of  value $M/R$  that is allowed before gravitational collapse occurs? (We will call this maximum value of $M/R$ the critical value of $M/R$). We have already mentioned that  for neutral perfect fluid spheres that the stability limit is given by the Buchdahl  limit \cite{Buchdahl}:
\begin{equation}
\left(\frac{M}{R}\right)  \leq \frac{4}{9}  {~~~~~\rm for~a~ perfect ~ fluid}.
\end{equation}
In  charged spheres the critical value of $M/R$ becomes dependent on the total charge $Q$, since the  addition of charge to the system  increases its total  energy and hence its total mass.   Andr\'{e}asson \cite{Andreasson}, 
has published a remarkable result that claims that for any compact spherically symmetric charged distribution  the following relationship holds between $M/R$ and $Q/R$:
\begin{equation}
\label{Andf}
\frac{M}{R} \leq \left(\frac{1}{3} + \sqrt{  \frac{1}{9} + \frac{1}{3} \frac{Q^2}{R^2} } ~\right)^2 {~~~\rm for ~all~charged~ spheres}.
\end{equation}
This formula generalizes the Buchdahl  result for perfect fluid neutral spheres. It is worth noting that in his derivation, Buchdahl placed the following constraints on the physical properties of the fluid: (i) $d\rho/dr  < 0$ i.e., the density  decreases outward from the centre of the sphere  and (ii) the pressure is isotropic.  In the derivation of his formula, Andr\'{e}asson made the following assumptions about  the matter content of the sphere: (i) $ p_r + 2 p_t < \rho$ 
 and (ii) $\rho >0$. 
We note that unlike the neutral case    Andr\'{e}asson     did not require the condition $d\rho/dr <0$ for the derivation of his formula.
There have been several numerical investigations of the upper bound of $M/R$ as a function of $Q/R$  and they have all verified that the  Andr\'{e}asson    formula provides a valid upper bound of $M/R$ for charged spheres  \cite{Lemos} and \cite{Lemos2}.   

In our previous paper \cite{Dev1} we showed that if the condition $d\rho/dr  > 0$ is allowed in charged perfect fluid spheres then the Andr\'{e}asson  limit can be violated. One of our aims in this study is to continue the investigation of the  the validity of the  Andr\'{e}asson  limit as it applies to charged anisotropic spheres using a mixture of analytical and numerical techniques. 


This paper is organized as follows in the next section we will briefly review the derivation of Einstein-Maxwell equations for the interior gravitational field of a charged anisotropic sphere. In section 3 we solve the field equations and consider their stability properties and the formation of extremal black holes in these solutions.  In section 4 discuss our results and formulate our conclusions. We note that a prime ($'$) or a comma ($,$) denotes derivatives with respect $r$, and a semi-colon ($;$) is used to represent covariant derivatives. We will work in units where $c = G = 1$.

\section{The field equations}
We are  interested in studying the interior gravitational field of  charged spheres, therefore we will assume a spherically symmetric metric of the form
\begin{equation}
ds^2  = - e^{2 \nu} dt^2 + e^{2 \lambda} dr^2 + r^2 d\theta^2 + r^2 \sin^2\theta d\phi^2.
\end{equation}
In this paper we are concerned only with  spherically symmetric static solutions  of the coupled Einstein - Maxwell equations, therefore $\nu \, = \, \nu(r)$ and $\lambda\,=\, \lambda(r)$ are functions of $r$ only.  
The Einstein field equations are 
\begin{equation}
\label{EF} 
G_{\alpha \beta} = R_{\alpha \beta} - \frac{1}{2} g_{\alpha \beta} R = 8 \pi T_{\alpha \beta}.
\end{equation}
The spheres that we propose to study have  a matter content that consists of  an anisotropic fluid with a charge distribution that gives rise to static radial  electric field. The energy-momentum tensor $T_{\alpha \beta}$,  will thus written as
\begin{equation}
T_{\alpha \beta}  = (T_{\alpha \beta})_{af} + (T_{\alpha \beta})_{em},
\end{equation}
with $(T_{\alpha  \beta})_{af}$  the energy-momentum tensor for an anisotropic fluid and  $  (T_{\alpha \beta})_{em}  $ the energy-momentum tensor   associated with the electric field. 

The most general energy momentum tensor  for a static spherically symmetric anisotropic fluid is
\begin{equation}
\label{Tfl}
(T_{\alpha \beta})_{af} = ( \rho + p_t) u_{\alpha} u_{\beta} + p_t g_{\alpha \beta}  +(p_r - p_t) v_{\alpha} v_{\beta}, 
\end{equation} 
where  $\rho, \, p_r $ and $p_t$ are the density, the radial pressure and the tangential pressure of the fluid respectively and are functions  of $r$ only,  $u^{\alpha}$ is the 4-velocity of the fluid and $v^{\alpha} $ is a unit space-like vector in the radial direction.   

The  details pertaining to the construction of the energy-momentum tensors and be found in \cite{Dev1}, here we will quote the result:
\begin{equation}
(T_{\alpha \beta })_ {af}= {\rm{diag}} (-\rho e^{-2\nu} , p_r \,e^{-2\lambda} , p_t \,r^2 , r^2\,  p_t \, \sin^2 \theta  ).
\end{equation} 
The electromagnetic energy-momentum tensor has the from 

\begin{equation}
(T_{\alpha \beta})_{em}  = \frac{q(r) ^2}{8 \pi r^4} {\rm diag} (e^{-2 \nu}, - e^{-2\lambda}, r^2, r^2 \sin^2 \theta).
\end{equation}
where
\begin{equation}
\label{charge}
q(r) = \int \limits_V  j^t  \sqrt{-g} \,dV,
\end{equation}
is the total charge contained in a  sphere of radius and $j^t$ is the time component of the four-current density.

The complete set of equations  that describe  the interior gravitational field of static anisotropic charged spheres are:
\begin{equation}
\label{gt}
G^t_{~t} = e^{-2 \lambda}  \left( \frac{2 \lambda^{\prime}} {r}  - \frac{1}{r^2} \right)  +  \frac{1}{r^2}     = 8 \pi \rho + \frac{q^2}{r^4} ,
\end{equation}

\begin{equation}
\label{grr}
G^r_{~r} = e^{-2 \lambda}  \left( \frac{2 \nu^{\prime}} {r}  + \frac{1}{r^2} \right)  -  \frac{1}{r^2}     = 8 \pi p_r - \frac{q^2}{r^4} ,
\end{equation}
and
\begin{equation}
\label{gth}
 G^\theta_{~\theta} = G^\phi_{~\phi }= e^{-2 \lambda}  \left(   \nu^{\prime \prime} +  {\nu^{\prime}}^2  - \nu^{\prime}\lambda^{\prime} + \frac{\nu^{\prime}} {r}  - \frac{\lambda^{\prime}}{r} \right)   
    = 8 \pi p_t + \frac{q^2}{r^4} .
\end{equation}

\section{Solutions for the field equations}
We will now develop solutions for the field equations. We start by noting that (\ref{gt}),  can be written as

\begin{equation}
\frac{d}{dr} (r e^{-2 \lambda} ) = 1 - 8 \pi \rho  r^2  - \frac{q^2}{r^2}.  
\end{equation}
This equation can be immediately integrated to give 
\begin{equation}
\label{eli}
e^{-2 \lambda(r)} = 1- \frac{2m_{i}(r)}{r} - \frac{f(r)}{r},
\end{equation}
with
\begin{equation}
m_{i}(r) =  4 \pi \int_0^{r}  \rho (s)  {s}^2 ds  ~~~{\rm and}~~~ f(r) = \int_0^r \frac{q(s)^2}{{s}^2}ds.
\end{equation}
The quantity $m_i(r) $ is the inertial mass of the  fluid in a sphere of radius $r$.  The requirement  that $e^{- 2 \lambda(r) }$ matches  the Reissner-Nordstr\"{o}m metric  at surface of the sphere, $r = R$ gives 
\begin{equation}
\label{mass}
1 - \frac{M}{R} + \frac{Q^2}{R^2}  = 1  - \frac{1}{R} \int_0^R ( 8 \pi \rho r^2  + \frac{q^2}{r^2} ) dr,
\end{equation}
where $M$ is the total mass and $Q$ is the total charge. An expression for the total mass can be found from this equation:
\begin{equation}
M =  \frac{1}{2} \int_0^R ( 8 \pi \rho r^2  + \frac{q^2}{r^2} ) dr  +  \frac{Q^2}{2R}.
\end{equation}
We note that $e^{-2\lambda(r)}$ can also be written as 
\begin{equation}
 \label{elg} 
e^{-2 \lambda(r)} = 1- \frac{2m_{g}(r)}{r} + \frac{q^2(r)}{r^2}.
\end{equation} 
where the function $q(r)$ is the total charge in a sphere of radius $r$. It is the  charge function defined in (\ref{charge}) . In  order for  (\ref{eli}) to be equal to (\ref{elg}), $m_g(r)$ must be defined in the following manner:
\begin{equation}
m_g(r) \equiv \frac{1}{2 } \int_0 ^r \left(8 \pi \rho(s) s^2 + \frac{q^2(s)}{s^2} \right)  d s + \frac{q(r)^2}{2 r}.
\end{equation}
$m_g(r)$ is the total gravitating mass in a sphere of radius $r$. At the surface of the sphere $m_g(r) $ is equal $M$ the total mass. 
We note that in the absence of the electric field $m_g = m_i$.

We will study charged spheres with the following density, $\rho(r)$ and charge, $q(r)$ profiles:
\begin{equation}
\label{rhoa}
\rho(r) = \rho_o - \frac{b  }{8 \pi}  \frac{Q^2}{R^6} r^2  {\rm ~~~~and~~~~}  q(r) =   \frac{Q}{R^3} r^3,
\end{equation}
in the expression for $\rho(r)$, $b$ is a number. In our models 
\begin{equation}
\label{elambda}
e^{-2 \lambda(r)} = 1- \frac{8 \pi \rho_o}{3} r^2  + \frac{( b  -1)}{5}  \frac{Q^2}{R^6} {r^4}
\end{equation}
\begin{equation}
\label{mgeq}
m_g(r)  =  \frac{4 \pi \rho_o}{3}r^3 +   \frac{(6 - b )}{10}\frac{Q^2}{R} \frac{r^5}{R^5}.
\end{equation}
and
\begin{equation}
\label{Meq}
M =  \frac{4 \pi \rho_o}{3}R^3 +   \frac{(6 - b )}{10}\frac{Q^2}{ R}.
\end{equation}
With the assumed fluid and charge density profiles here,  when $b = 6$ 
\begin{equation}
 \frac{m_g(r)}{r^3} = \frac{M}{R^3} = const,
\end{equation}
thus, the $b = 6$ model is a sphere with a constant gravitational mass density. Also all  models have 
\begin{equation}
\frac{ q(r) }{r^3} =  \left( \frac{Q}{R^3}\right)  = const.
 \end{equation}
 thus, the charge density is constant for all our models. 
We will study in detail three charged configurations:
\begin{enumerate}
\item  $b = 0$  - a sphere with a constant  density fluid and constant charge density. 
\item  $b = 1$ -  a sphere with constant total energy density and constant charge density.
\item $b = 6 $ - a sphere with constant gravitational mass density and constant charge density. 
\end{enumerate}
We can solve for $\rho_o$ from (\ref{Meq}) and rewrite (\ref{elambda}) as
\begin{equation}
\label{lambda}
e^{-2 \lambda(r)} = 1-  \left( \frac{2M}{R}  + \frac{(b-6)}{5}  \frac{Q^2}{R^2}    \right) \frac{r^2}{R^2}  + \frac{(b-1)}{5}  \frac{Q^2}{R^2} \frac{r^4}{R^4}.
\end{equation}
Thus, if we  are given $\rho(r) $ and $q(r)$ we have found $ \lambda(r)$. 

We now need to solve for $\nu(r)$. We start by transforming (\ref{gth}). First we subtract   (\ref{grr})  from  (\ref{gth} )  to get
\begin{equation}
\fl  e^{-2 \lambda}  \left(   \nu^{\prime \prime} +  {\nu^{\prime}}^2  - \nu^{\prime}\lambda^{\prime} - \frac{\nu^{\prime}} {r}   \right)  
 -  \frac{\lambda^{\prime} e^{-2 \lambda}}{r}                         -  \frac{ e^{-2 \lambda}}{r^2}  +  \frac{1}{r^2}                                                                              = 2 \frac{q^2}{r^4}  + 8 \pi (p_t - p_r).
\end{equation}
Then we substitute for $1/r^2$ from   (\ref{gt})   to get  the following equation
\begin{equation}
\label{w1}
  \nu^{\prime \prime} +  {\nu^{\prime}}^2  - \nu^{\prime}\lambda^{\prime} - \frac{\nu^{\prime}} {r}   
= 3 \frac{ \lambda^{\prime}} {r}  -  \left( 8 \pi \rho  - \frac{q^2}{r^4}  - 8\pi (p_t - p_r) \right)   e^{2\lambda} .
 \end{equation}
We next multiply both sides of this equation  by $e^{- \lambda + \nu}/r$, and we find that the left hand-side becomes an exact differential: $ ((\nu^{\prime} e^{- \lambda + \nu})/r)^{\prime}$. Introducing $\zeta(r)\equiv e^{\nu(r)}$, we can write (\ref{w1}) in the following form 
\begin{equation}
\label{w2}
\left( \frac{1}{r} e^{-\lambda} \zeta^{\prime} \right)^{\prime}  = \left[ \frac{3 \lambda^{\prime} e^{-2 \lambda}} {r^2}  -   \frac{8 \pi \rho}{r}  + \frac{q^2}{r^5}  +  \frac{8\pi }{r}(p_t - p_r)  \right] e^{\lambda} \zeta.
\end{equation} 

A similar equation was derived by Giuliani and Rothman \cite {Giuliani} in their study of charged perfect fluid spheres.  The transformation of the left hand-side  (\ref{w1}) into  that of (\ref{w2}) was introduced by Weinberg \cite{Weinberg} in deriving the equation he used to prove the Buchdahl limit of $M/R$ for neutral perfect fluid spheres.  

We need to define the form of the anisotropy in order to solve \ref{w2}. Here we propose that the anisotropy is proportional to the electromagnetic energy density:
\begin{equation}
 8\pi (p_t - p_r)  = j \frac{q^2}{r^4}  = j\frac{Q^2}{R^6}r^2,
 \end{equation} 
 with  $j$ is a constant. 
Substituting this form of the anisotropy,  $\rho(r)$ and  $q(r)$ from (\ref{rhoa}) and $\lambda^{\prime}e^{- 2 \lambda}$ from (\ref{lambda})  we find that (\ref{w2}) becomes 
\begin{equation}
\label{zeta}
\left( \frac{1}{r} e^{-\lambda} \zeta^{\prime} \right)^{\prime}  = \left( \frac{11}{5} - \frac{b}{5}  + j\right) \frac{Q^2}{R^6} r e^{\lambda} \zeta.
\end{equation}
We now define a new variable 
\begin{equation}
\label{utransform}
\tilde{\zeta}( u(r))  \equiv \zeta(r)  {\rm ~~~~~with~~~~}  u(r)  = \frac{1}{R^2} \int_0^r s e^{\lambda(s) } ds.
\end{equation}
Then (\ref{zeta}) becomes
\begin{equation}
\label{Th}
\frac{ d^2 \tilde{\zeta}}{d u^2}  = a \frac{Q^2}{R^2}  \tilde{\zeta} (u) . 
\end{equation}
with  $( 11/5 -b/5  + j )\,=\,  a$. This is our master equation. Our task now is to find solutions for it given values of $b$ and $j$. There are three types of solutions of (\ref{Th}) depending on the value of $a$ :
 \begin{equation} 
 \label{a1eq}
 \fl (i) ~~a    > 0    :  ~~ \tilde{\zeta} (u(r)) = A e^{\sqrt{a} \frac{Q}{R} u(r)}  +  B  e^{-\sqrt{a} \frac{Q}{R} u(r)}   
\end{equation}
\begin{equation}
 \label{a2eq}
 \fl (ii) ~~   a  = 0              :  ~~\tilde{\zeta} (u(r))   = A  +  B  u(r) ,
\end{equation}
\begin{equation}
 \label{a33eq}
 \fl  (iii)~~   a  < 0           : ~~  \tilde{\zeta} (u(r)) = A \sin\left(\sqrt{|a|} \frac{Q}{R} u(r)\right)  +  B  \cos \left(\sqrt{|a|} \frac{Q}{R} u(r)\right)  \nonumber \\
\end{equation}

 
 
The values of the constants $A$ and $B$ in (\ref{a1eq}), (\ref{a2eq})  and (\ref{a33eq}) are fixed by the the boundary conditions imposed of $\tilde{\zeta}(u)$:
\begin{equation}
\tilde{\zeta}(u(R)) =  \left(1 - \frac{2M}{R} + \frac{Q^2}{R^2}\right)^{\frac{1}{2}} 
\end{equation}
and
\begin{equation}
\frac{d\tilde{\zeta}(u(R))}{du(R) }  =  \left(\frac{M}{R} - \frac{Q^2}{R^2}\right).
\end{equation}
The condition for stability is $\zeta(r = 0) = 0$. The critical values of $M/R$ as a function of $Q/R$ are found from solving the equation $\zeta(r=0) = 0$. We will call the $\zeta(r=0) = 0$ equation the critical values equation.

The introduction of anisotropy into the field equations results in an extra degree of freedom in the equation that determines $\zeta(r)$.  In case of perfect fluids, a given value of $b$ uniquely determines $a$ in the master equation and this gives a one to one correspondence between $b$ and $\zeta(r)$. In the anisotropic case, the presence of the free parameter $j$ in the expression for $a$  allows us to generate a very large number of values for $a$ for a given value $b$. Thus for each $b$ there are now a very large number of $\zeta(r)$'s. It is clearly not possible to study all of the solutions, therefore here we will restrict ourselves to studying the changes that occur when anisotropy is added to some of the perfect fluid models that we studied in \cite{Dev1}. In particular, we will study three models in detail here: $b = 0$ (a fluid with constant inertial density plus constant charge density), $b = 1$ (a fluid with constant total energy density and constant charge density) and $b = 6$ (a fluid with constant gravitational mass density and constant charge density). 

\subsection {Solutions with  b = 0. }
When $b = 0$,
\begin{equation}
e^{-2 \lambda(r)} = 1-  \left( \frac{2M}{R}  - \frac{6}{5} \frac{Q^2}{R^2}    \right) \frac{r^2}{R^2}  -  \frac{1}{5}  \frac{Q^2}{R^2} \frac{r^4}{R^4},
\end{equation}

\begin{eqnarray}
\fl u(r) =    \frac{1}{R^2} \int_0^r   e^{\lambda(s)}  s ds              
= \frac{\sqrt{5}}{2}\frac{R}{Q} \left[ \tan^{-1}  \left(  \left(            {\sqrt{5}}  \frac{M}{Q} -  \frac{3}{\sqrt{5}}\frac{Q}{ R }+  \frac{1}{\sqrt{5}}  \frac{Q }{R^3} r^2 \right) e^{\lambda(r)} \right)  \right. \\ \nonumber
~~~~~~~~~~~~~~~~~~~~~~~~~~~~~~~~~~~~~~~~~~~~~~\left.  -\tan^{-1}       \left(  {\sqrt{5}}  \frac{M}{Q} -  \frac{3}{\sqrt{5}}\frac{Q}{ R }\right)\right],
\end{eqnarray}
and the master equation becomes 
\begin{equation}
\frac{ d^2 \tilde{\zeta}}{d u^2}  = \left(\frac{11}{5}  + j \right)  \frac{Q^2}{R^2}  \tilde{\zeta} (u) . 
\end{equation}
There are three distinct  types solutions  here, solutions with $j > -11/5$,  $j = -11/5$ and $j < - 11/5$. We will consider the $j > - 11/5$ case first. 
\subsubsection{Solutions with  b = 0, and  j > -11/5.}
\hspace{0.2in}

\begin{figure}
\centering
  \includegraphics[width=0.6\linewidth]{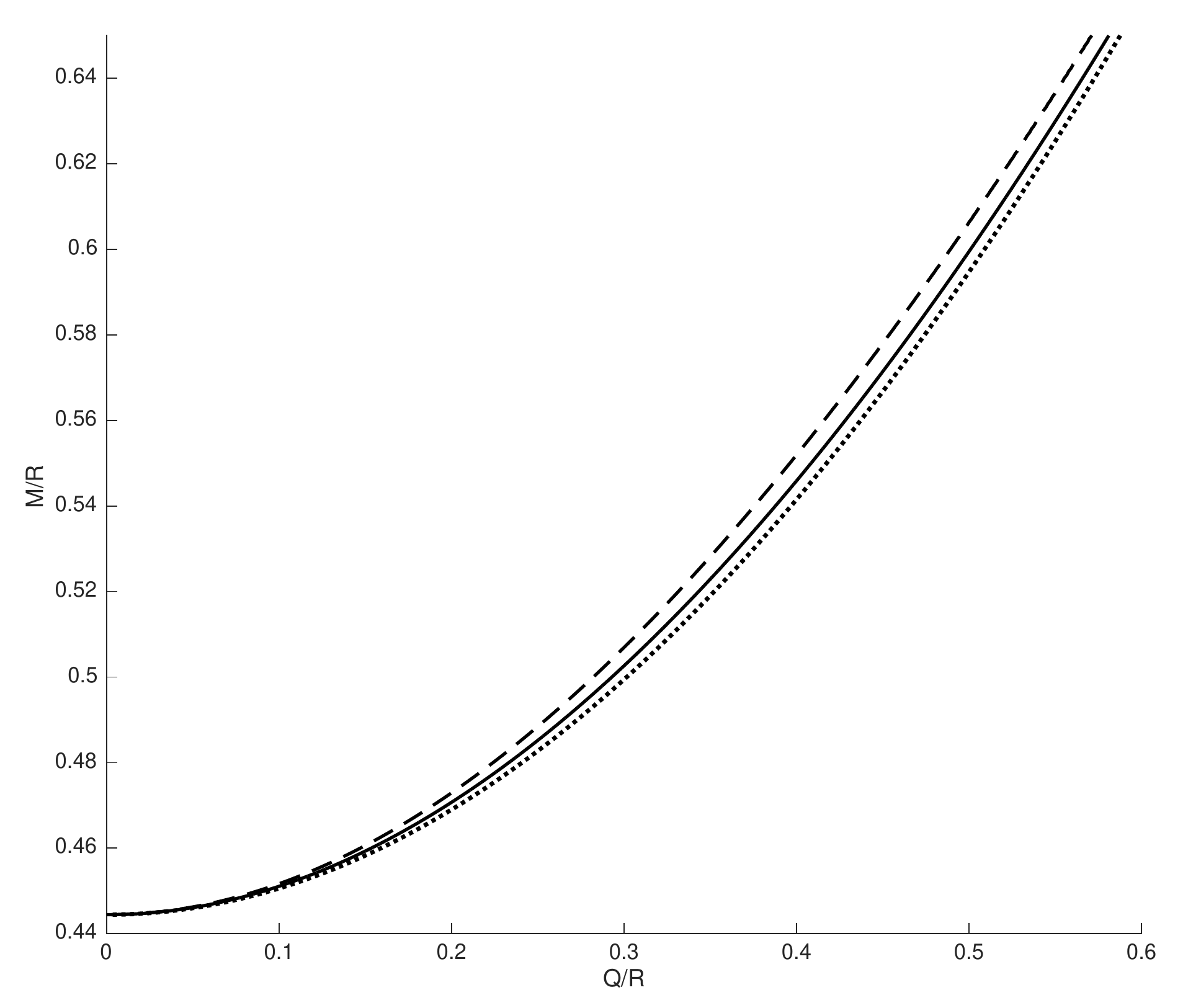}
  \caption{    The critical values of M/R vs Q/R from (\ref{anj1}) for $b = 0$ and $j = 8$   ({\bf - - - - -}), the  Andr\'{e}asson formula   (\rule{0.9 cm}{0.03 cm})  and  the isotropic model  ({\bf $\cdot$$\cdot$$\cdot$$\cdot$$\cdot$$\cdot$$\cdot$}) . 
        }
  \label{fig:f1}
\end{figure}

\noindent For $ b = 0,$ and $ j > -11/5$, the master equation is 
\begin{equation}
\frac{ d^2 \tilde{\zeta}}{d u^2}  =  k^2  \frac{Q^2}{R^2}  \tilde{\zeta} (u) ~~~~~~{\rm with} ~~ k = \left(\frac{11}{5} +  j\right)^{\frac{1}{2}}
\end{equation}
The solutions here are 
\begin{equation}
\zeta(r) =    A \exp \left[{ k \frac{Q}{R} u(r)} \right]             + B  \exp \left[{ -k\frac{Q}{R} u(r)} \right]  
\end{equation}
The constants $A$ and $B$ are found using the boundary conditions. Solving  for them we find we find, 
\begin{equation}
 A = \frac{1}{2} \left[ \left(1 - \frac{2M}{R} + \frac{Q^2}{R^2}\right)^{\frac{1}{2}}  +  \frac{1}{k}       \left(\frac{M}{Q} - \frac{Q}{R}\right)  \right] \exp \left[ -k \frac{Q}{R} u(R) \right]   
\end{equation}
and
\begin{equation}
B = \frac{1}{2} \left[ \left(1 - \frac{2M}{R} + \frac{Q^2}{R^2}\right)^  {\frac{1}{2}}     -       \frac{1}{k}      \left(\frac{M}{Q} - \frac{Q}{R}\right) \right] \exp \left[ k \frac{Q}{R} u(R) \right]   
\end{equation}
The stability condition is $\tilde{\zeta} (u(r =0 ) = 0) $. Here  $\tilde{\zeta} (u(r =0 ) = 0) $ requires $A + B$  = 0. This leads to the following  critical values equation:
\begin{eqnarray}
\label{anj1}
\fl{   \left[      \frac{1}{k}       \left(\frac{M}{Q} - \frac{Q}{R}\right)      +            \left(1 - \frac{2M}{R} + \frac{Q^2}{R^2}\right)^{\frac{1}{2}}  \right] }=   \left[     \frac{1}{k}       \left(\frac{M}{Q} - \frac{Q}{R}\right)      -    
    \left(1 - \frac{2M}{R} + \frac{Q^2}{R^2}\right)^{\frac{1}{2}}  \right]   \\ \nonumber
\fl ~~~~~~~\exp \left[ \sqrt{5} k \left[       \tan^{-1} \left(  \left(  {\sqrt{5}}   \frac{M}{Q}  - \frac{2}{ \sqrt{5}}\frac{Q}{ R } \right) e^{\lambda(R)}
 \right)   -\tan^{-1}   \left(   {\sqrt{5}}   \frac{M}{Q}  - \frac{3}{ \sqrt{5}}\frac{Q}{ R }                                \right)                                                    \right] \right].
 \end{eqnarray}
 We studied this equation numerically and  found that when $j > 3 $ the critical values of $M/R$ are greater than the corresponding values of $M/R$ from the      Andr\'{e}asson formula for a given $Q/R$.  Figure \ref{fig:f1}  show that the critical values here for   $j = 8$ are greater  than the corresponding values from   Andr\'{e}asson formula and the isotropic model.  


\subsubsection{Solution for  b = 0  and  j = -11/5.}
\hspace{0.2in}

\noindent When  $ b = 0 $ and $ j = -11/5$ the master equation becomes 
\begin{equation}
\frac{ d^2 \tilde{\zeta}}{d u^2}  = 0
\end{equation}
The solution for $\zeta(r)$ here is 
\begin{eqnarray}
\fl {\zeta} (r)   = \frac{\sqrt{5}}{2} \left(\frac{M}{Q} - \frac{Q}{R} \right) \left[ \tan^{-1}  \left(  \left(            \sqrt{5}  \frac{M}{Q} -  \frac{3}{\sqrt{5}}\frac{Q}{ R }+  \frac{1}{\sqrt{5} } \frac{Q }{R^3} r^2 \right) e^{\lambda(r)} 
\right)  \right. \\ \nonumber
~~~~~~~~~~~~~\left.  -\tan^{-1}  \left(     \left(  {\sqrt{5}}  \frac{M}{Q} - \frac{2}{\sqrt{5}}\frac{Q}{ R }\right) e^{\lambda(R)}\right)\right]  + \left(1 - \frac{2M}{R} + \frac{Q^2}{R^2}\right)^{\frac{1}{2}}
\end{eqnarray}
The stability condition $\zeta( r = 0 ) \, = \,0$ gives the following  implicit equation for the dependence of the critical values of $M/R$ on $Q/R$: 
\begin{eqnarray}
\label{anj2}
\fl ~~~~~\left(1 - \frac{2M}{R} + \frac{Q^2}{R^2}\right)^{\frac{1}{2}}  = \frac{\sqrt{5}}{2 } \left(\frac{M}{Q} - \frac{Q}{R} \right) \left[ 
 \left(  \tan^{-1}  \left(     \left(  {\sqrt{5}}  \frac{M}{Q} - \frac{2}{\sqrt{5}}\frac{Q}{ R }\right) e^{\lambda(R)}\right) \right. \right.
  \\ \nonumber
~~~~~~~~~~~~~~~~~~~~~~~~~~~~~~~~~~~~~~~~~~\left.   -\tan^{-1}  \left(             {\sqrt{5}}  \frac{M}{Q} -  \frac{3}{\sqrt{5}}\frac{Q}{ R } \right) \right].
 \end{eqnarray}
 
 
 \begin{figure}
\centering
 \includegraphics[width=0.6\linewidth]{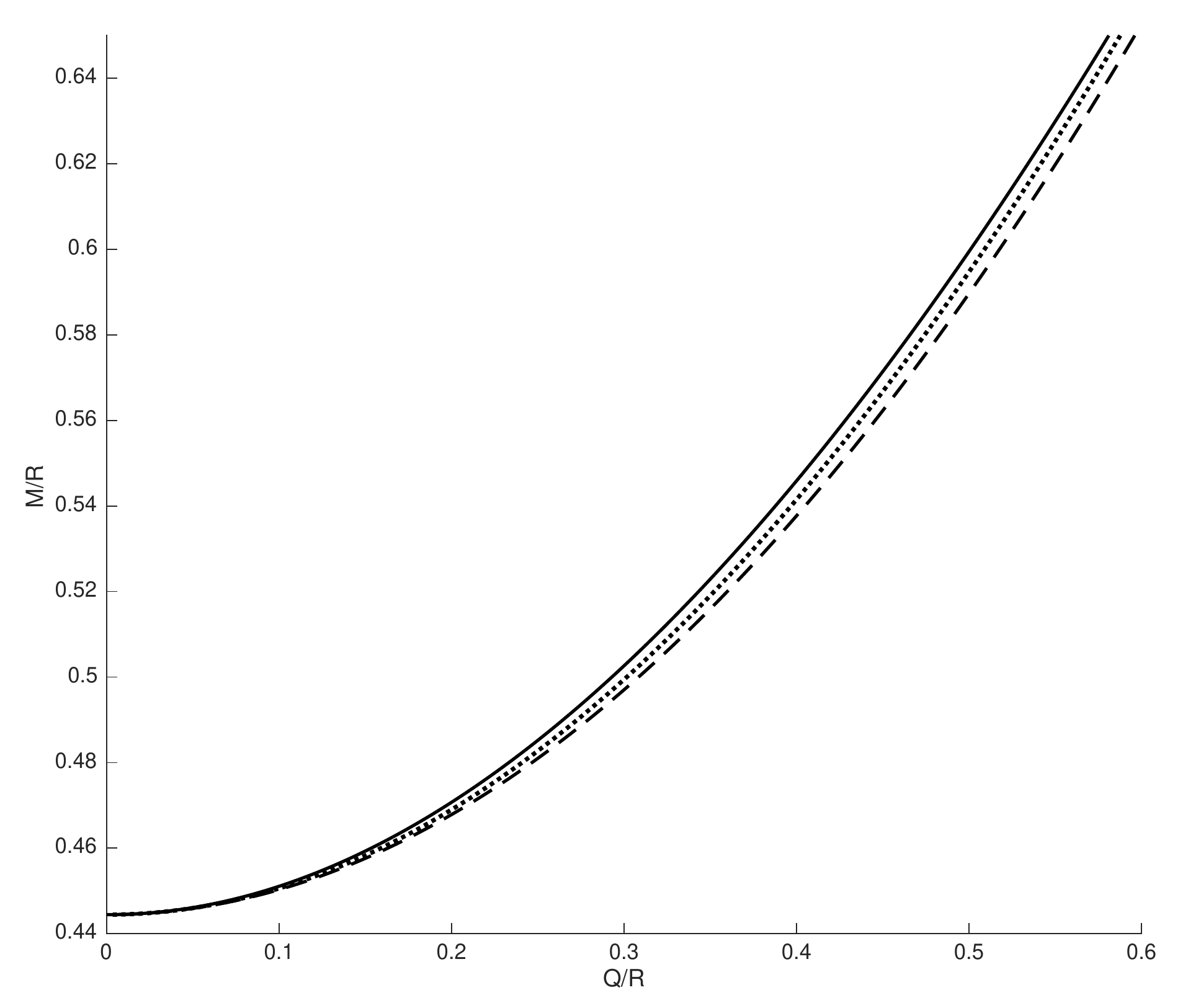}
 \caption{    The critical values of M/R vs Q/R from (\ref{anj2}) for $b = 0$ and $j = -11/5 $  ({\bf - - - - -}), the  Andr\'{e}asson formula   (\rule{0.9 cm}{0.03 cm})  and  the isotropic model  ({\bf $\cdot$$\cdot$$\cdot$$\cdot$$\cdot$$\cdot$$\cdot$}) .  } 
  \label{fig:f2}
\end{figure}

This equation was solved numerically and the results were plotted in Figure \ref{fig:f2}. We observe that the critical values for this model  are less  than those from the the  Andr\'{e}asson formula and the $j = 0$  isotropic model.


\subsubsection{Solutions with  b = 0, and  j < -11/5.}
\hspace{0.2in}

\noindent When  $ b = 0$ and $j < -11/5$, the master equation becomes 
\begin{equation}
\label{Th00}
\frac{ d^2 \tilde{\zeta}}{d u^2}  = -d^2 \frac{Q^2}{R^2}  \tilde{\zeta} (u), ~~~{\rm with} ~~~ d =  \sqrt{|11/5 +j|}
\end{equation}
The solution for $\tilde{\zeta}(u(r)$ here is 
\begin{equation}
\tilde{\zeta}(u(r)) = A \sin \left(d \frac{Q}{R} u(r) \right)  + B \cos \left(d  \frac{Q}{R} u(r) \right) 
\end{equation}
with 
 \begin{equation}
 A =     e^{-\lambda(R)}        \sin\left(d\frac{Q}{R} u(R) \right )         +  \frac{1}{d} \left(\frac{M}{Q} - \frac{Q}{R} \right) \cos\left(d\frac{Q}{R} u(R) \right )  ,
\end{equation}

\begin{equation}
  B =     e^{-\lambda(R)}     \cos\left(d\frac{Q}{R} u(R) \right )        - \frac{1}{d} \left(\frac{M}{Q} - \frac{Q}{R} \right) \sin\left(d\frac{Q}{R} u(R) \right ).
\end{equation}

\begin{figure}
\centering
 \includegraphics[width=0.6\linewidth]{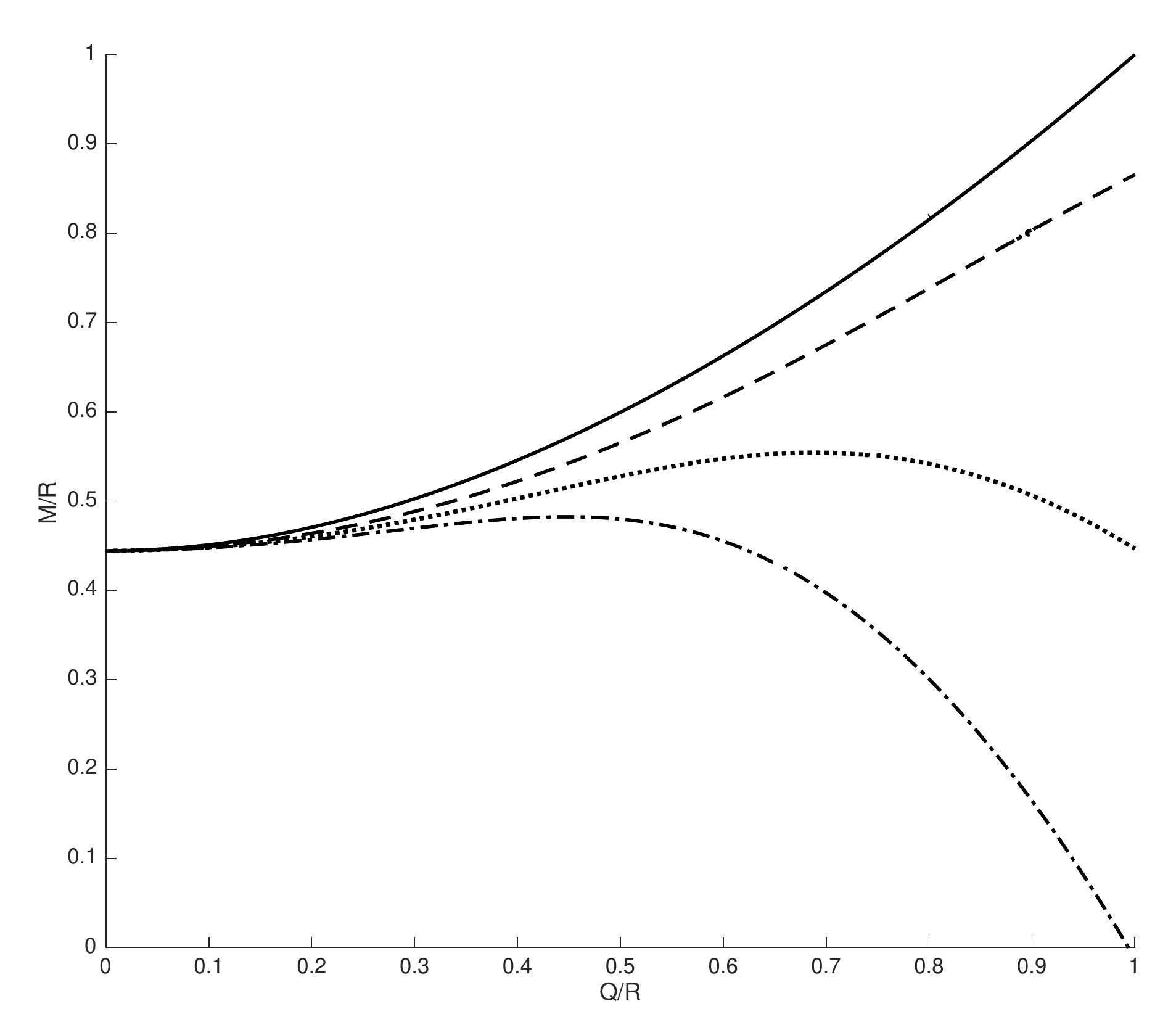}
  \caption{  The critical values of M/R vs Q/R  for $b = 0$ and  $j = -9$      $({\bf- - - -})$,     $j = -15$   $({\bf \cdots \bf \cdots        }) $               and    $j = -20.3$    $({\bf \cdot-\cdot-\cdot})$  respectively    from (\ref{aneq3})    and     the  Andr\'{e}asson formula   (\rule{0.9 cm}{0.03 cm}). }
  \label{fig:f3}
\end{figure}

\noindent The stability condition $\zeta(r= 0) = 0$ requires
\begin{equation}
 A  \sin\left(d\frac{Q}{R} u(0) \right )           +  B \cos \left( d\frac{Q}{R} u(0)  \right) = 0,
 \end{equation} 
however, since $u(0) \equiv 0$, then  $B$ must be equal to zero here. This results in   the following critical values equation:
\begin{equation}
 e^{-\lambda(R)}     \cos\left(d\frac{Q}{R} u(R) \right )        = \frac{1}{d} \left(\frac{M}{Q} - \frac{Q}{R} \right) \sin\left(d\frac{Q}{R} u(R) \right ) 
 \end{equation}
Here 
\begin{eqnarray}
\fl u(R) =   
 \frac{\sqrt{5}}{2}\frac{R}{Q} \left[ \tan^{-1}  \left(  \left(            {\sqrt{5}}  \frac{M}{Q} -  \frac{2}{\sqrt{5}}\frac{Q}{ R } \right) e^{\lambda(R)} \right)   -\tan^{-1}       \left(  {\sqrt{5}}  \frac{M}{Q} -  \frac{3}{\sqrt{5}}\frac{Q}{ R }\right)\right],
\end{eqnarray}
thus critical values of $M/R \, \,vs \,\, Q/R$ are solutions of the following equation:
\begin{eqnarray}
\label{aneq3}
\fl \left( 1 - \frac{2 M}{R}  + \frac{Q^2}{R^2} \right)^{\frac{1}{2}}  = \frac{1}{d}\left(\frac{M}{Q} - \frac{Q}{R} \right) \tan \left[  \frac{\sqrt{5}}{2} d  \left[ \tan^{-1}  \left(  \left(            {\sqrt{5}}  \frac{M}{Q} -  \frac{2}{\sqrt{5}}\frac{Q}{ R } \right) e^{\lambda(R)} \right) \right.  \right. \\ \nonumber 
~~~~~~~~~~~~~~~~~~~~~~~~~~~~~~~~~~~~~~~~~~~~ -\tan^{-1} \left. \left.      \left(  {\sqrt{5}}  \frac{M}{Q} -  \frac{3}{\sqrt{5}}\frac{Q}{ R }\right)\right]\right].
\end{eqnarray} 
This equation has an exact solution for $d = \sqrt{4/5}$  i.e., $j = -3$.  The expression is 
\begin{equation}
\frac{M}{R} = \frac{1}{2} \left[   1  + \frac{Q^2}{R^2}  - \left( {\tilde{Q}}       +  \frac{16}{81\tilde{Q}  }                    -     \frac{Q^2}{15 \tilde{Q} R^2}   - \frac{5}{9}  \right)^{2}  \right]
\end{equation}
with 
\begin{equation}
\fl ~~~~~~~~~~~~~\tilde{Q} = \left[ \frac{64}{729}  - \frac{Q^2}{9 R^2}  + \left(\frac{1}{3375} \frac{Q^6}{R^6} + \frac{59}{6075} \frac{Q^4}{R^4}  - \frac{128}{10935} \frac{Q^2}{R^2} \right)^{\frac{1}{2}}   \right]^{\frac{1}{3}}
\end{equation} 
We solved (\ref{aneq3}) numerically for various values of $d = \sqrt{ |11/5 + j|}$ and some results are plotted in Figure \ref{fig:f3}. We found that for $-9  < j < -3$ the curves are similar to the curve from  the   Andr\'{e}asson formula   but the values of $M/R$ from these curves  are less that those from the   Andr\'{e}asson formula. Also the $M/R$ values for a given $Q/R$ decreases with decreasing $j$. When $j \approx -10$ the critical values curves start to turn downward and away from the Andr\'{e}asson formula curve and when $j \approx -20.3$ the critical value of $M/R$ for $Q/R = 1$ is zero. This behavior  of the critical values curves with increasing negative values of $j$ is  similar to the isotropic case with increasing positive values of $b$. 

\begin{figure}[h!]
  \centering
  \begin{subfigure}[b]{0.45\linewidth}
    \includegraphics[width=\linewidth]{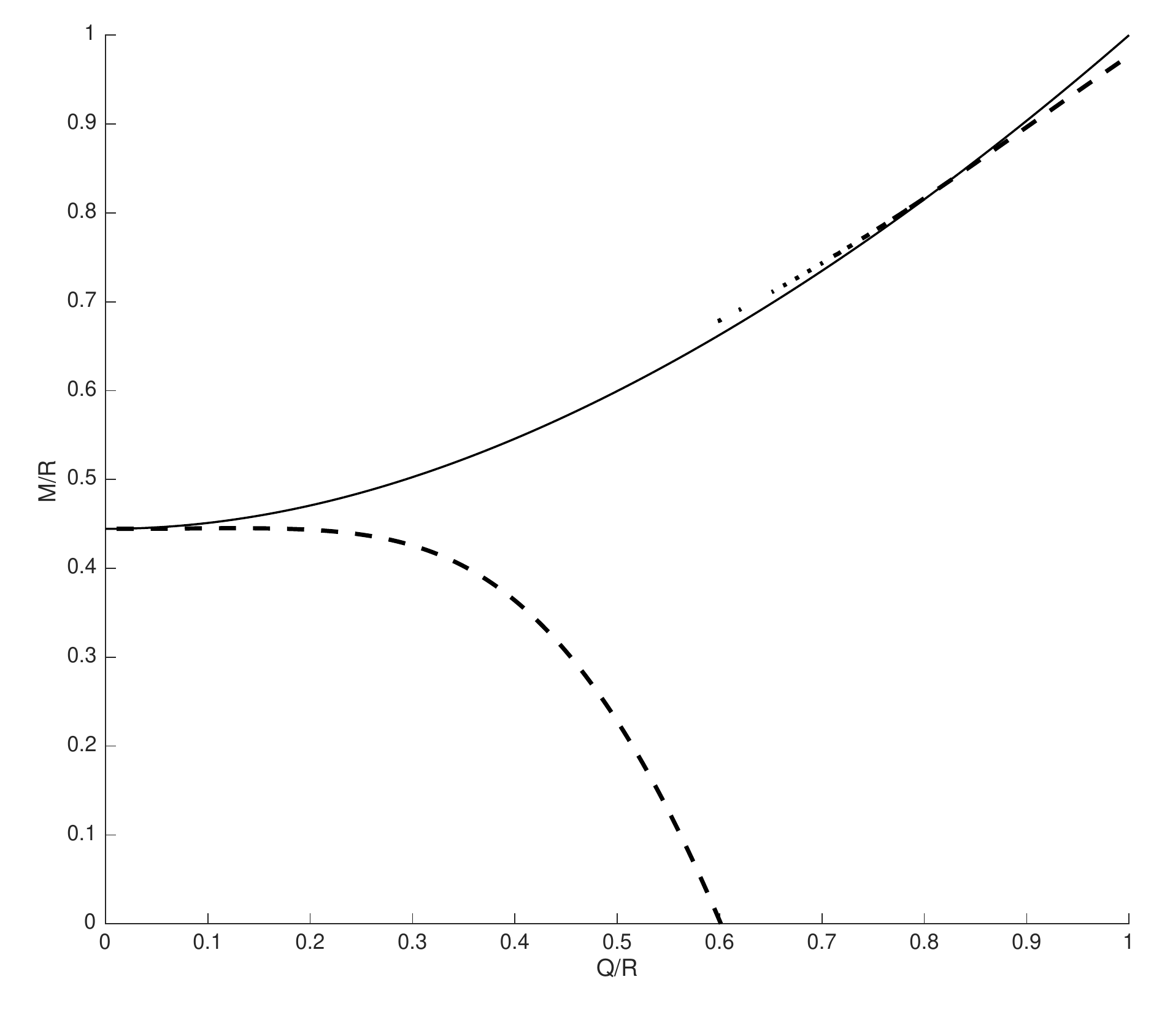}
     \caption{ $d = 6$.}
  \end{subfigure}
  \begin{subfigure}[b]{0.45\linewidth}
    \includegraphics[width=\linewidth]{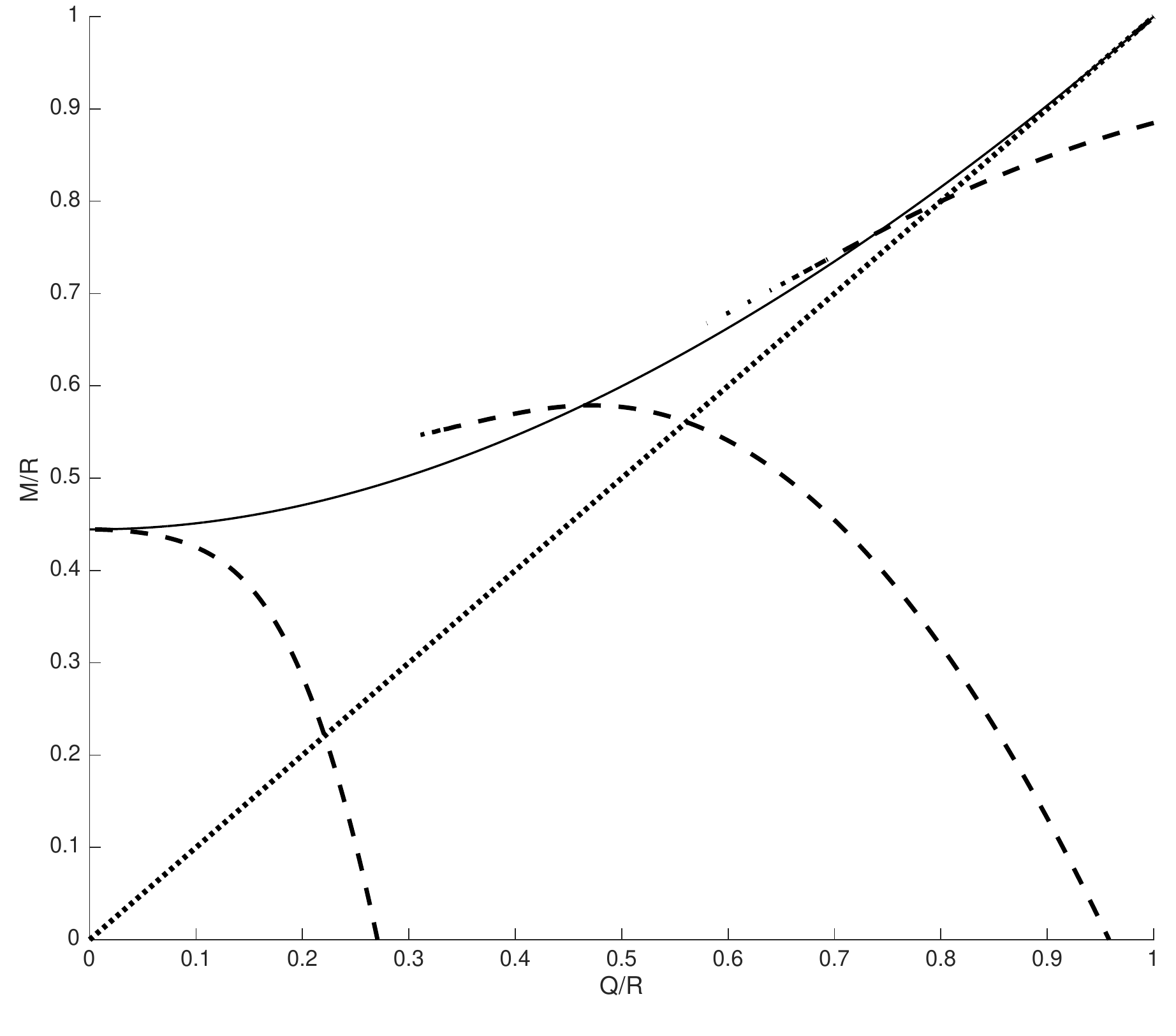}
    \caption{$d = 12.$}
  \end{subfigure}
  \begin{subfigure}[b]{0.45\linewidth}
    \includegraphics[width=\linewidth]{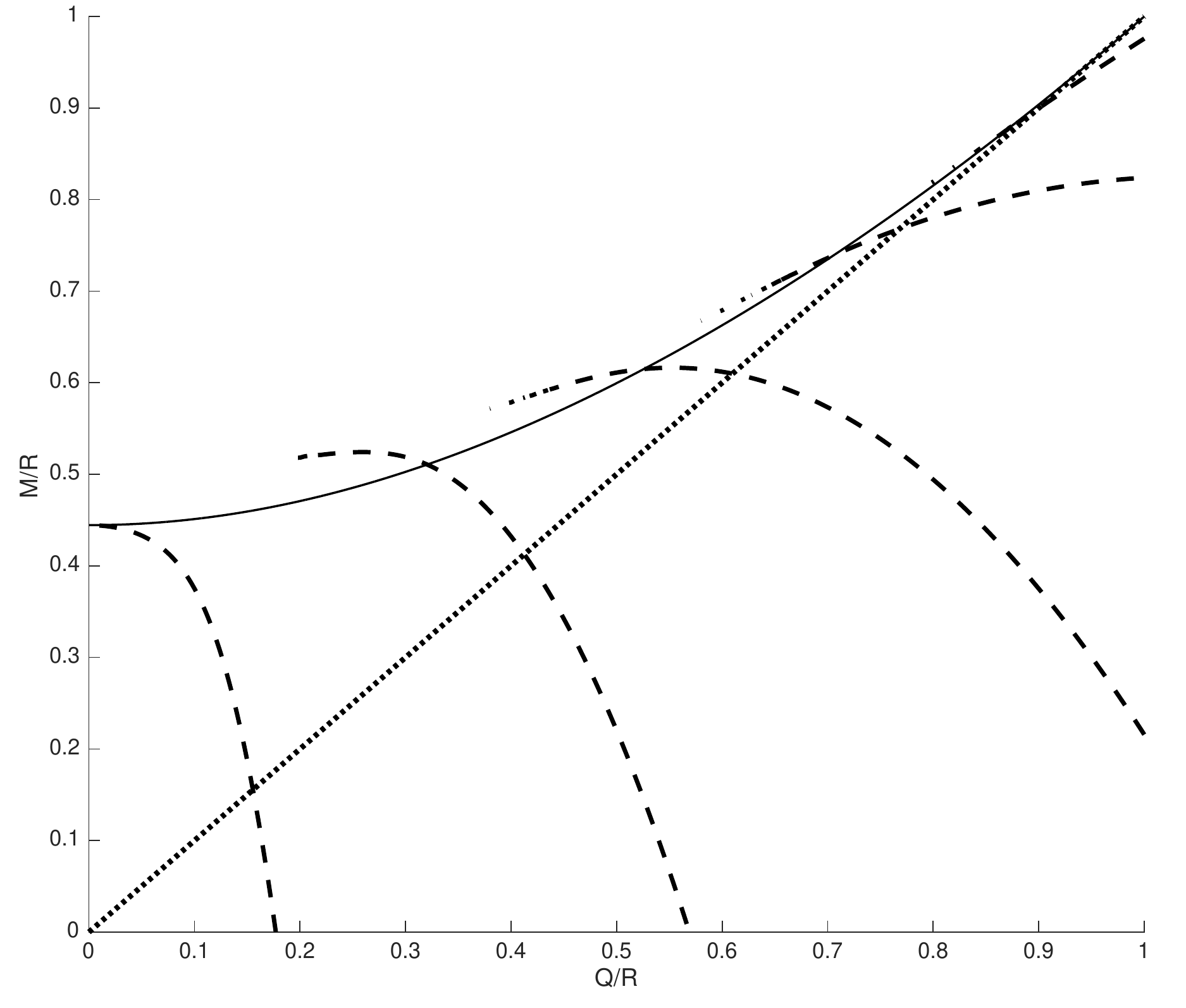}
    \caption{$d = 18.$}
  \end{subfigure}
  \begin{subfigure}[b]{0.45\linewidth}
    \includegraphics[width=\linewidth]{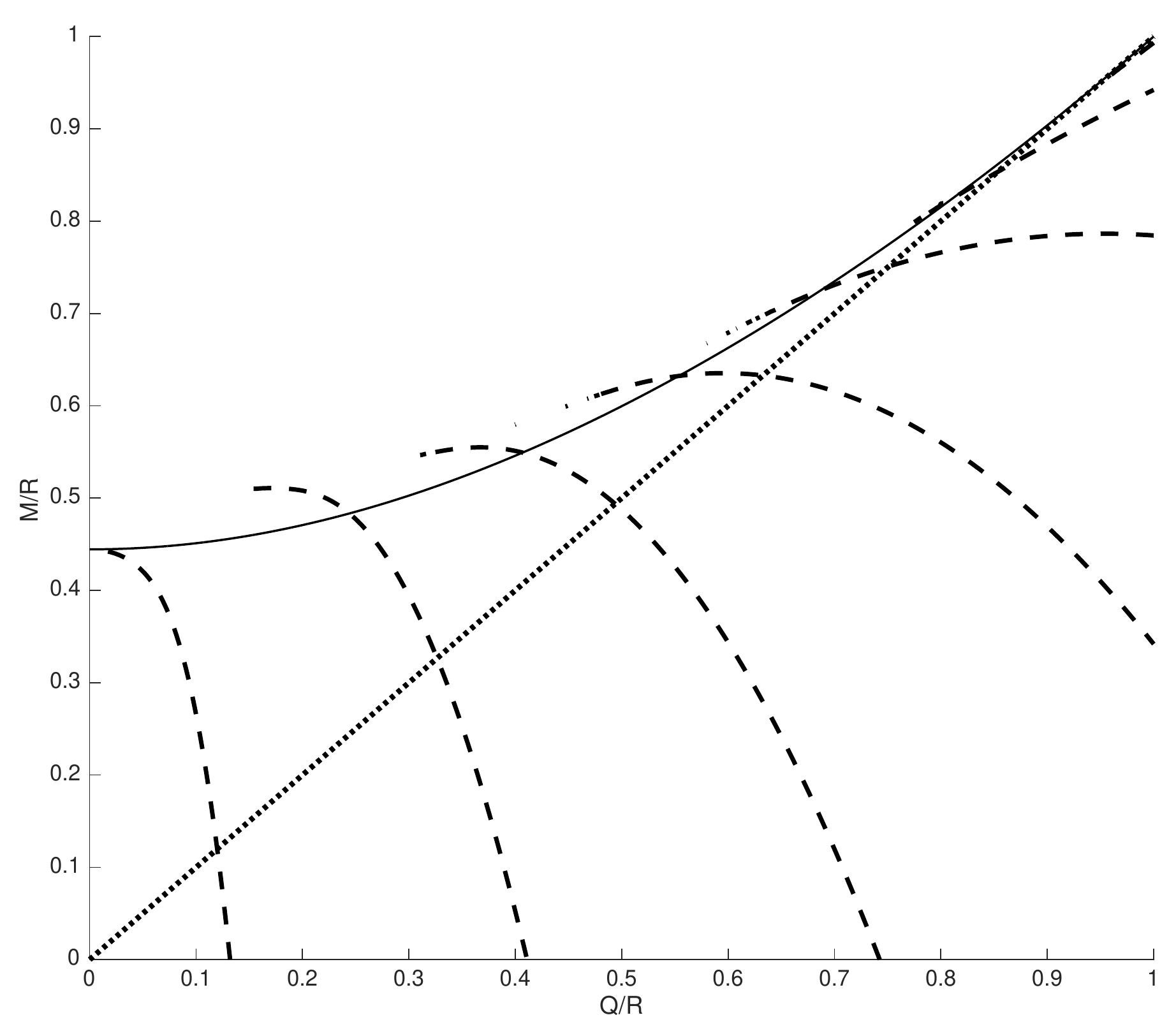}
    \caption{$ d = 24$.}
  \end{subfigure}
  \caption{ \it The critical values curves (${\bf ----}$) that are obtained from (\ref{aneq3})  by varying $d$. The dotted line   $({\bf \cdot \cdot \cdot \cdot \cdot})$     is the  $M/R = Q/R$ extremal line     and     the curve from  Andr\'{e}asson formula is the solid line  (\rule{0.9 cm}{0.03 cm}). }
  \label{fig:f20}
\end{figure}

Here, for large  negative $j$ we found a new feature in the critical values curves that was not present in the isotropic models. For large negative $j$  (or positive $d = \sqrt{|11/5 + j|}$)  the critical values curves  instead of being  continuous  now becomes disjointed and the segments have a quasi-periodic behavior. This quasi-periodic behavior results in there being several extremal values for each $j$. This behavior is plotted in Figure \ref{fig:f20}.


\subsection{ Solutions with b = 1}
 When $ b = 1$,  
 \begin{equation}
e^{-2 \lambda(r)} = 1-  \left( \frac{2M}{R}  -   \frac{Q^2}{R^2}    \right) \frac{r^2}{R^2} ,
\end{equation}

\begin{eqnarray}
\fl~~~~~~~~~~~~u(r) =  \frac{1}{R^2} \int_0^r   e^{\lambda(s)}  s ds  =  \left( \frac{2M}{R} - \frac{Q^2}{R^2} \right)^{-1}  \left[ 1 -  \left[  1-  \left( \frac{2M}{R}  -   \frac{Q^2}{R^2}    \right) \frac{r^2}{R^2}\right]^{\frac{1}{2}} \right].
\end{eqnarray}
and the master equation becomes 
\begin{equation}
\label{k12}
\frac{ d^2 \tilde{\zeta}}{d u^2}  = ( 2  + j ) \frac{Q^2}{R^2}  \tilde{\zeta} (u).
\end{equation} 
The solutions are divided into three types: (i) $j > -2$, $j = -2$ and $j > -2$.

\subsubsection{ Solutions for b = 1 and j > -2.}
\hspace{0.2in}

\begin{figure}
\centering
  \includegraphics[width=0.6\linewidth]{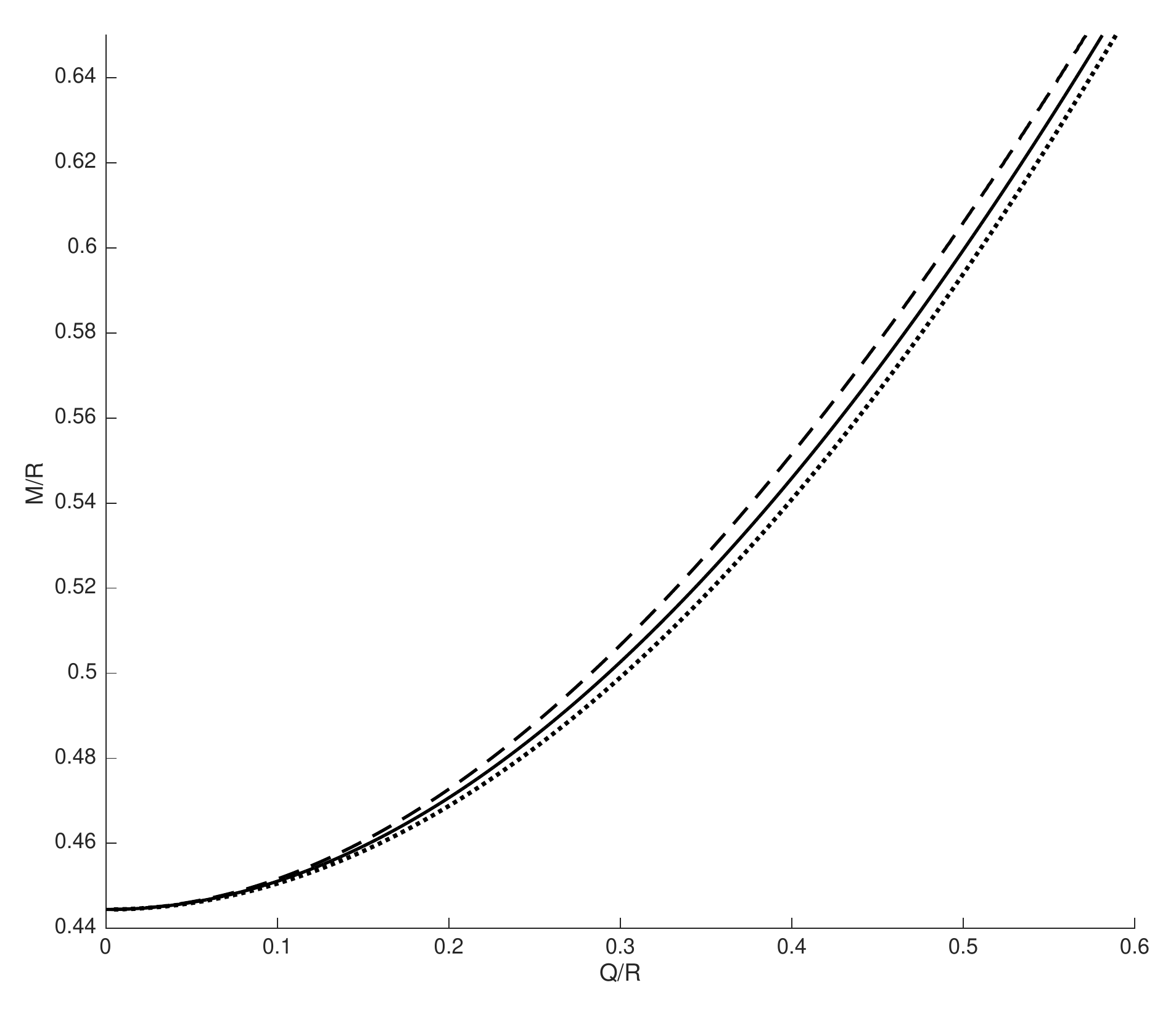}
  \caption{    The critical values of M/R vs Q/R from (\ref{aneq5}) for  $j = 8$   ({\bf - - - - -}), the  Andr\'{e}asson formula   (\rule{0.9 cm}{0.03 cm})  and  the isotropic model  ({\bf $\cdot$$\cdot$$\cdot$$\cdot$$\cdot$$\cdot$$\cdot$}) . 
        }
  \label{fig:f4}
\end{figure}

The solution for (\ref{k12}) with $j  > - 2$   is 
\begin{equation}
\fl ~~~~~~~~~~~\tilde{\zeta} (u(r))  = A  \exp \left(h \frac{Q}{R} u(r)  \right)          +  B \exp \left(h \frac{Q}{R} u(r)  \right) ~~{\rm with} ~~h = (2 +j)^{\frac{1}{2}}
 \end{equation}
 with
 \begin{equation}
A = \frac{1}{2} \left[ \left(1 - \frac{2M}{R} + \frac{Q^2}{R^2}\right)^{\frac{1}{2}}  + \frac{1}{h} \left(\frac{M}{Q} - \frac{Q}{R}\right) \right] \exp \left (-h \frac{Q}{R} u(R)  \right),
\end{equation}

\begin{equation}
B = \frac{1}{2} \left[ \left(1 - \frac{2M}{R} + \frac{Q^2}{R^2} \right)^{\frac{1}{2}}  - \frac{1}{h} \left(\frac{M}{Q} - \frac{Q}{R}\right) \right] \exp\left( h \frac{Q}{R} u(R)  \right)
\end{equation}
and
\begin{eqnarray}
u(R) 
=  \left( \frac{2M}{R} - \frac{Q^2}{R^2} \right)^{-1}  \left[ 1 -  \left(  1-  \frac{2M}{R}  +   \frac{Q^2}{R^2}    \right)^{\frac{1}{2}} \right].
\end{eqnarray}
Here the critical values of $M/R$ are found from the following equation:
\begin{eqnarray}
\label{aneq5}
\fl \left[          \left(\frac{M}{Q} - \frac{Q}{R}\right)           +      h \left(1 - \frac{2M}{R} + \frac{Q^2}{R^2}\right)^{\frac{1}{2}}               \right] = \\ \nonumber
\fl ~~~~~{ \left[        \left(\frac{M}{Q} - \frac{Q}{R}\right)           -      h \left(1 - \frac{2M}{R} + \frac{Q^2}{R^2}\right)^{\frac{1}{2}}           \right] }  
\exp \left [ \frac{2 h }{\left( \frac{2M}{Q} - \frac{Q}{R} \right)} 
 \left[ 1 -  \left(  1-  \frac{2M}{R}  +   \frac{Q^2}{R^2}    \right)^{\frac{1}{2}} \right] \right]
\end{eqnarray}
The numerical results from (\ref{aneq5}) shows that  critical values of $M/R$ for curves with   $j  > 4 $ exceeds the values from the
Andr\'{e}asson formula. Also the critical values of $M/R$ for a given $Q/R$ increases with increasing $j$. The critical values curve for $j = 8$, the Andr\'{e}asson formula and the isotropic $b = 1$ are shown in Figure \ref{fig:f4}. 



\subsubsection{ Solution for b = 1 and j = -2}
\hspace{0.2in}

\begin{figure}
\centering
  \includegraphics[width=0.6\linewidth]{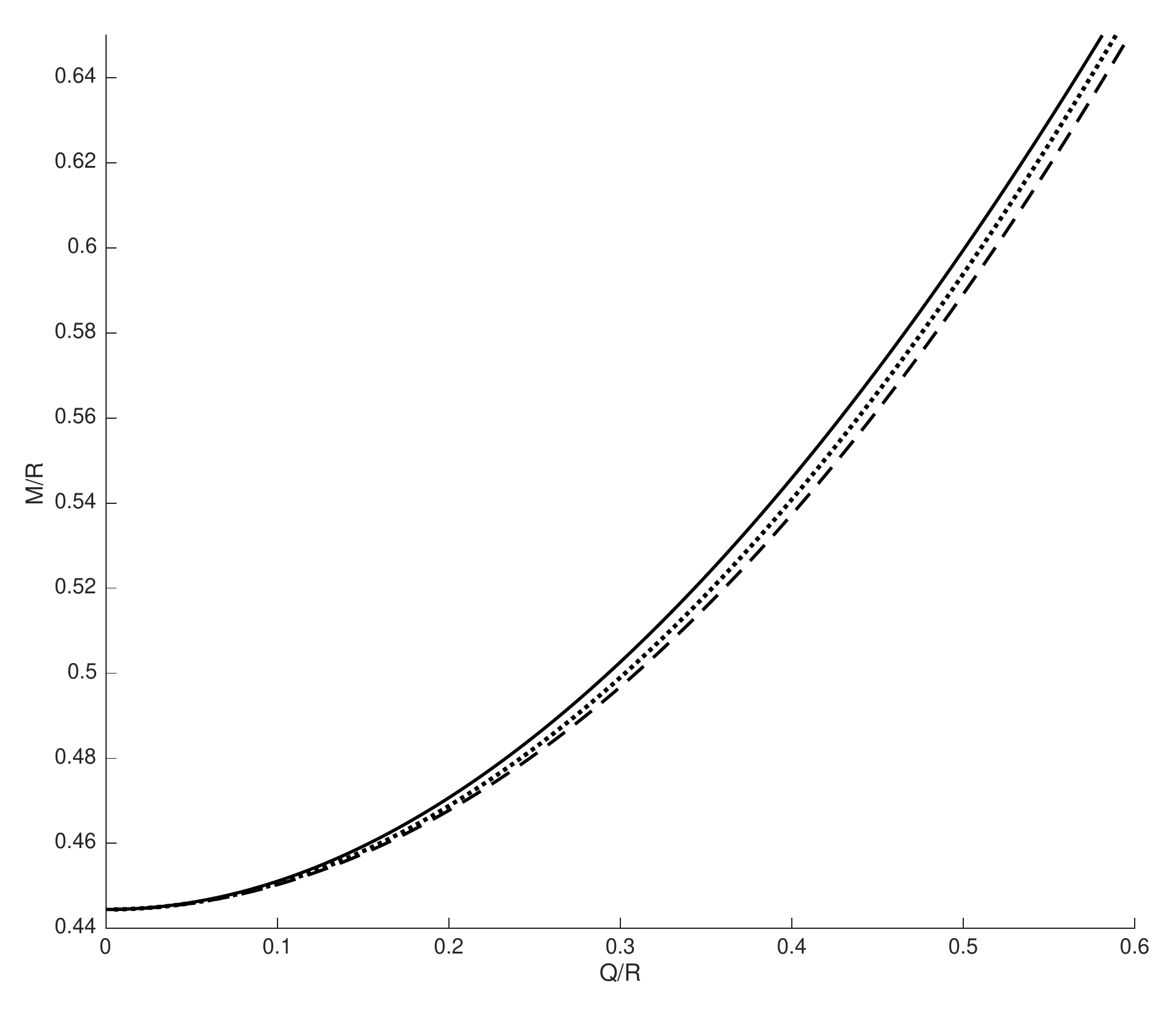}
  \caption{    The critical values of M/R vs Q/R  for the $b = 1$ and  $j = 2$  model  ({\bf - - - - -}), the  Andr\'{e}asson formula   (\rule{0.9 cm}{0.03 cm})  and  the isotropic model $b = 1$ model ({\bf $\cdot$$\cdot$$\cdot$$\cdot$$\cdot$$\cdot$$\cdot$}) . 
        }
  \label{fig:f5}
\end{figure}

When  $b = 1$ and $j = -2$ there is only one solution for the master equation. Here 
\begin{equation}
\frac{ d^2 \tilde{\zeta}}{d u^2}  = 0.
\end{equation} 
\noindent The solution for $\zeta(r)$ is 
\begin{eqnarray}
\fl~~~~~~~~~\zeta(r) =  \left(1 - \frac{2M}{R} + \frac{Q^2}{R^2}\right)^{\frac{1}{2}} +     \left(\frac{M}{R} - \frac{Q^2}{R^2}\right)   \left( \frac{2M}{R} - \frac{Q^2}{R^2} \right)^{-1}  \\ \nonumber
 ~~~~~~~~~~~~~~~~~~~~~~~~~~\left[  \left(1-  \frac{2M}{R}  -   \frac{Q^2}{R^2}  \right)^{\frac{1}{2}}-  \left(  1-  \left( \frac{2M}{R}  -   \frac{Q^2}{R^2}    \right) \frac{r^2}{R^2}\right)^{\frac{1}{2}}\right]
\end{eqnarray}
The critical values equation from the stability condition $\zeta(r =0) = 0$, has a very simple form here:
\begin{equation}
 \left(\frac{M}{R} - \frac{Q^2}{R^2}\right)  =  \left( \frac{3M}{R} - \frac{2Q^2}{R^2} \right) \left(1-  \frac{2M}{R}  -   \frac{Q^2}{R^2}  \right)^{\frac{1}{2}}.
 \end{equation}
 The roots of this equation are 
 
  \begin{equation}
  \label{aneq6}
 \fl ~~~~~~~~~~~~\frac{M}{R}  =  \left[ \frac{1}{2} \frac{Q^2}{R^2}, ~~~   \frac{2}{9} +  \frac{2}{3} \frac{Q^2}{R^2}  - \frac{1}{9}\left(4 - \frac{3Q^2}{R^2}\right)^{\frac{1}{2}}, ~~~\frac{2}{9} +  \frac{2}{3} \frac{Q^2}{R^2}  + \frac{1}{9}\left(4 - \frac{3Q^2}{R^2}\right)^{\frac{1}{2}} \right]
 \end{equation} 
  The first two roots require $M/R = 0$ when $Q/R = 0$. These solutions do not correspond to the properties of the model that we are studying here. Here, when $Q/R = 0$ we should have $M/R = 4/9$. The third root does give this result and thus we will take it as the solution for the critical values equation here. 
We note that it  is the same exact expression that we found  when studying the $b = 6$ isotropic model.  The critical	values of $M/R$ here are less the values from both the isotropic $b = 1$ model and the   Andr\'{e}asson formula. A plot is shown in Figure \ref{fig:f5}.

 \subsubsection{ Solutions for b = 1 and j  <-2}
\hspace{0.2 in}

When  $b = 1$ and $j < -2$ the master equation becomes 
\begin{equation}
\frac{ d^2 \tilde{\zeta}}{d u^2}  = - l^2  \frac{Q^2}{R^2}  \tilde{\zeta} (u) ~~~{\rm with} ~~ l = \sqrt{| 2 + j |}
\end{equation}  

The solution for $\tilde{\zeta}(u(r)$ here is 
\begin{equation}
\tilde{\zeta}(u(r)) = A \sin \left(l \frac{Q}{R} u(r) \right)  + B \cos \left(l  \frac{Q}{R} u(r) \right) 
\end{equation}
with 
 \begin{equation}
 A =     e^{-\lambda(R)}        \sin\left(l\frac{Q}{R} u(R) \right )         +  \frac{1}{l} \left(\frac{M}{Q} - \frac{Q}{R} \right) \cos\left(l\frac{Q}{R} u(R) \right )  ,
\end{equation}

\begin{equation}
  B =     e^{-\lambda(R)}     \cos\left(l\frac{Q}{R} u(R) \right )        - \frac{1}{l} \left(\frac{M}{Q} - \frac{Q}{R} \right) \sin\left(l\frac{Q}{R} u(R) \right ).
\end{equation}


\noindent The stability condition $\zeta(r= 0) = 0$ requires
\begin{equation}
 A  \sin\left(l\frac{Q}{R} u(0) \right )           +  B \cos \left( l\frac{Q}{R} u(0)  \right) = 0,
 \end{equation} 
however, since $u(0) \equiv 0$, then  $B$ must be equal to zero here. This results in   the following critical values equation:
\begin{equation}
 e^{-\lambda(R)}     \cos\left(l\frac{Q}{R} u(R) \right )        = \frac{1}{l} \left(\frac{M}{Q} - \frac{Q}{R} \right) \sin\left(l\frac{Q}{R} u(R) \right ) 
 \end{equation}
Here 
\begin{eqnarray}
u(R)   
=  \left( \frac{2M}{R} - \frac{Q^2}{R^2} \right)^{-1}  \left[ 1 -  \left(  1-  \frac{2M}{R}  +   \frac{Q^2}{R^2}    \right)^{\frac{1}{2}} \right].
\end{eqnarray}
thus critical values of $M/R \, \,vs \,\, Q/R$ are solutions of the following equation:
\begin{eqnarray}
\label{aneq6a}
\fl~~~~~~~~~~~~  \left( 1 - \frac{2 M}{R}  + \frac{Q^2}{R^2} \right)^{\frac{1}{2}}  = \frac{1}{l}\left(\frac{M}{Q} - \frac{Q}{R} \right) \tan \left[   \frac{ l\left[ 1 -  \left(  1-  \frac{2M}{R}  +   \frac{Q^2}{R^2}    \right)^{\frac{1}{2}} \right]   }{ \left( \frac{2M}{Q} - \frac{Q}{R} \right)  }         
\right].
\end{eqnarray} 

\begin{figure}
\centering
 \includegraphics[width=0.6\linewidth]{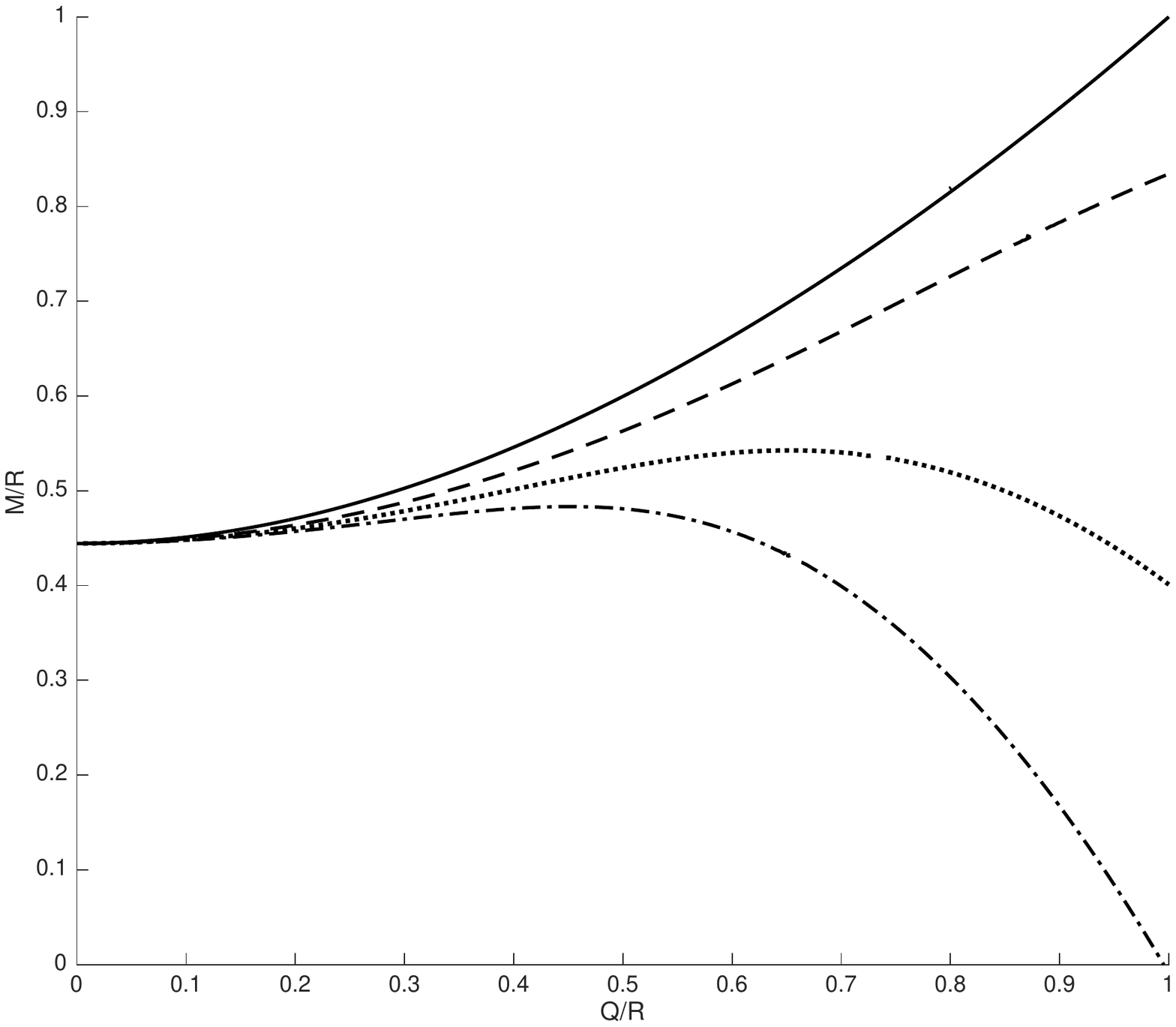}
  \caption{  The critical values of M/R vs Q/R  for  $j = -9$      $({\bf- - - -})$,     $j = -15$   $({\bf \cdots \bf \cdots        }) $               and    $j = -19.7$    $({\bf \cdot-\cdot-\cdot})$     from (\ref{aneq6a})    and     the  Andr\'{e}asson formula   (\rule{0.9 cm}{0.03 cm}). }
  \label{fig:f6}
\end{figure}

\begin{figure}[h!]
  \centering
  \begin{subfigure}[b]{0.45\linewidth}
    \includegraphics[width=\linewidth]{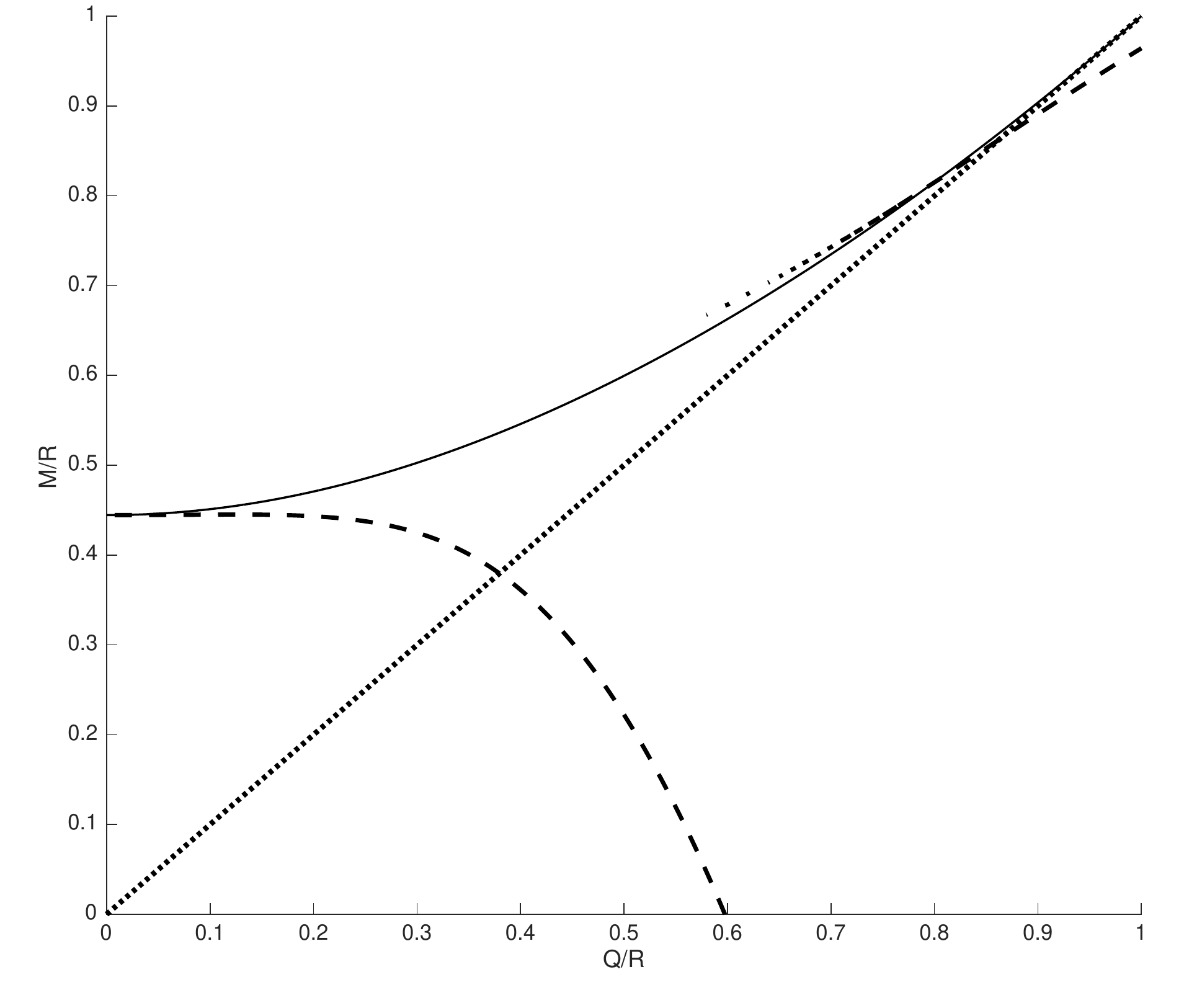}
     \caption{ $l = 6$.}
  \end{subfigure}
  \begin{subfigure}[b]{0.45\linewidth}
    \includegraphics[width=\linewidth]{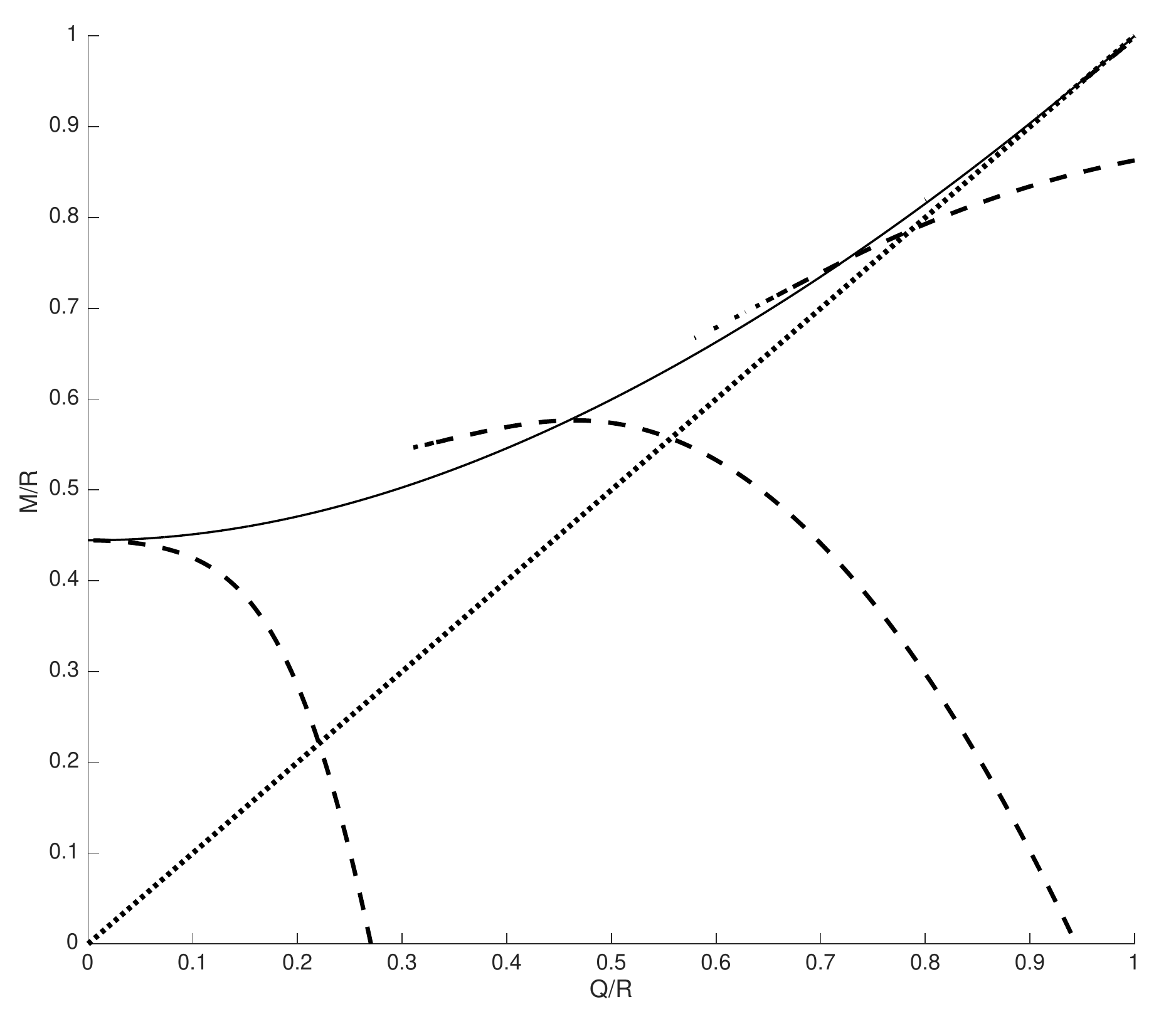}
    \caption{$l = 12.$}
  \end{subfigure}
  \begin{subfigure}[b]{0.45\linewidth}
    \includegraphics[width=\linewidth]{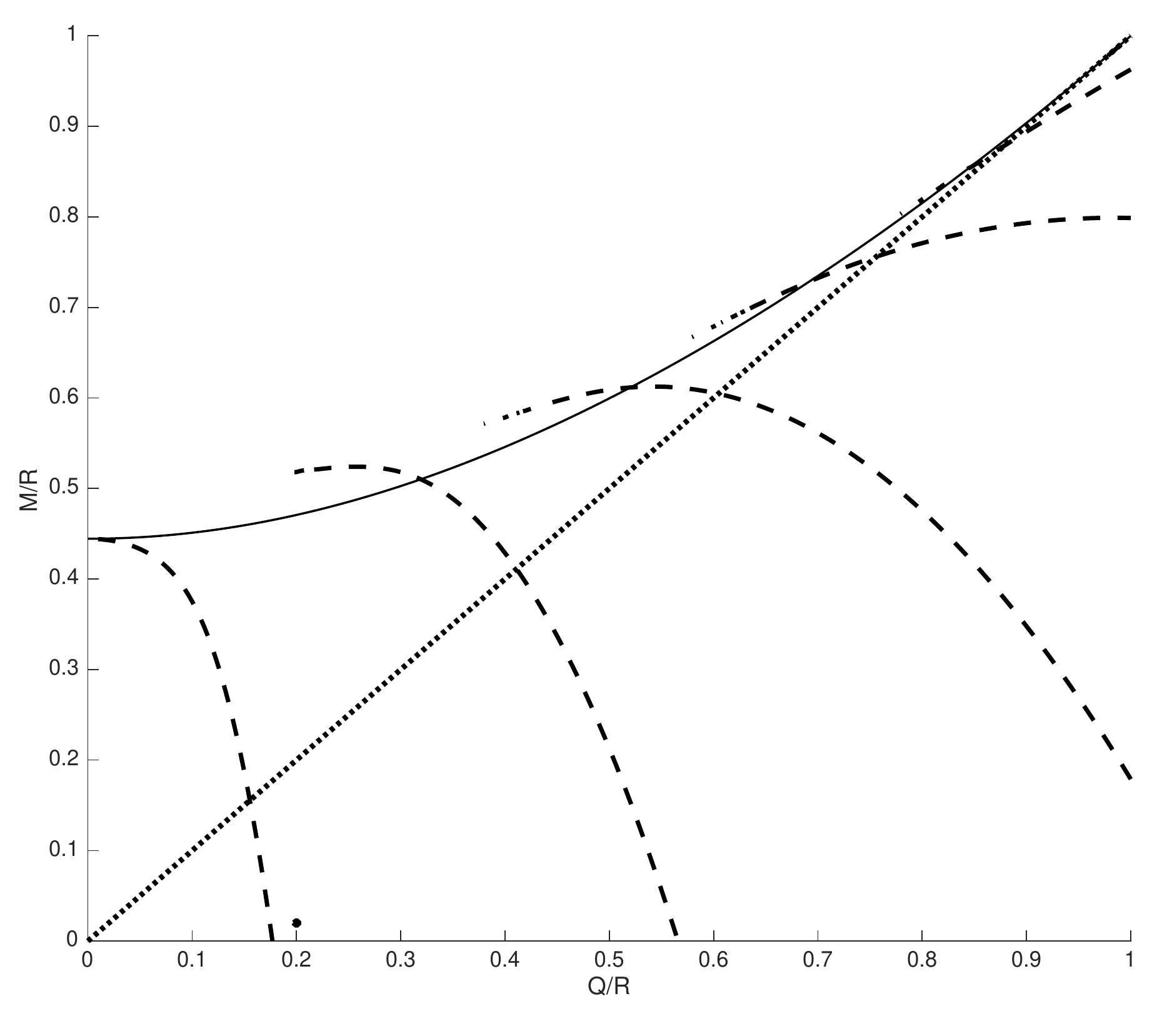}
    \caption{$l = 18.$}
  \end{subfigure}
  \begin{subfigure}[b]{0.45\linewidth}
    \includegraphics[width=\linewidth]{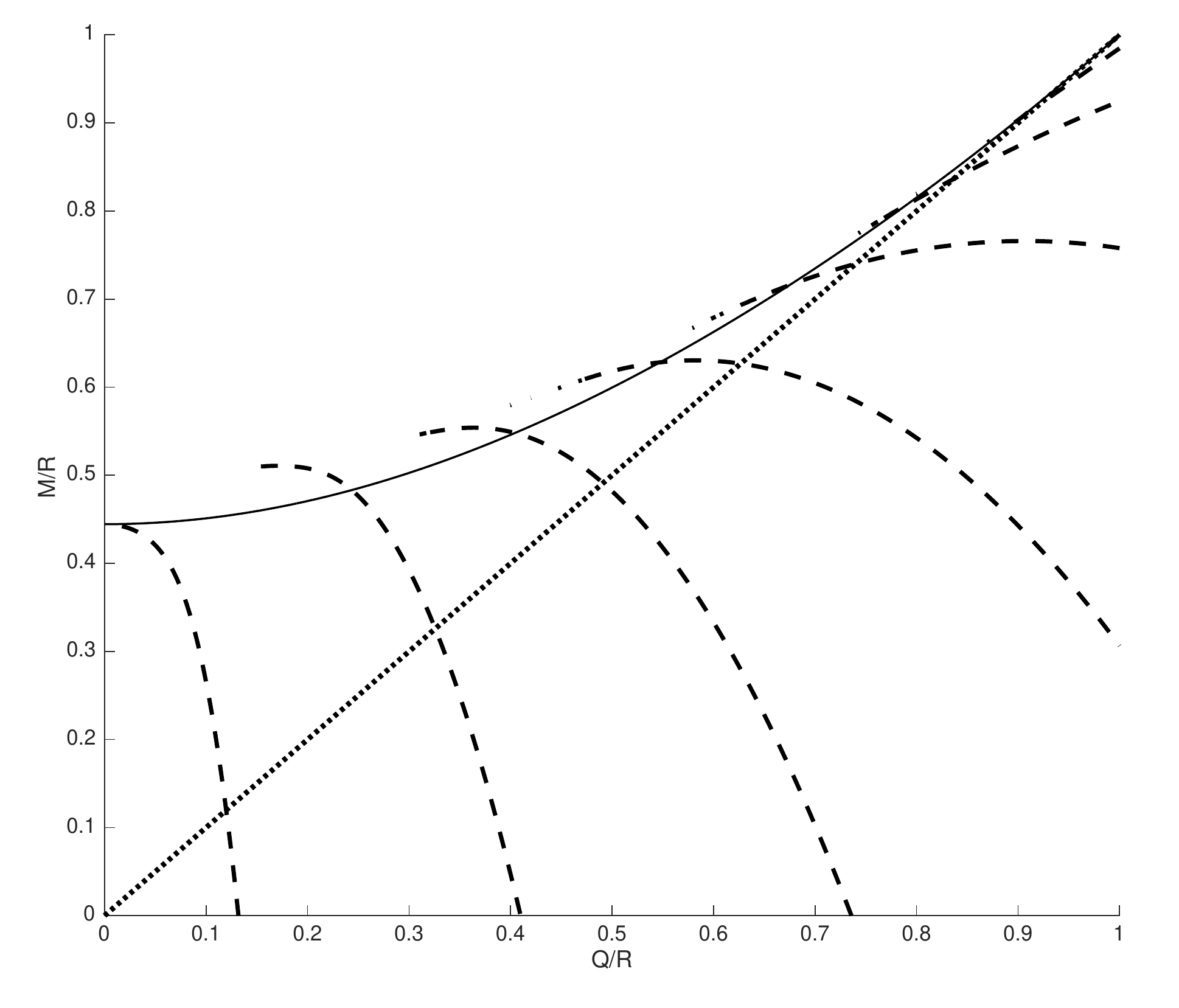}
    \caption{$ l = 24$.}
  \end{subfigure}
    \caption{ \it The critical values curves (${\bf ----}$) that are obtained from (\ref{aneq6a})  by varying $l$. The dotted line   $({\bf \cdot \cdot \cdot \cdot \cdot})$     is the  $M/R = Q/R$ extremal line     and     the curve from  Andr\'{e}asson formula is the solid line  (\rule{0.9 cm}{0.03 cm}). }
  \label{fig:f23}
\end{figure}

We solved (\ref{aneq6a}) numerically for various values of $l = \sqrt{ |2 + j|}$ and some results are plotted in Figure \ref{fig:f6}. We found  that the curves are similar to the $b = 0$ model, however for a given $j$ here the values of $M/R$ are less  than the $b = 0$ model and  $j \approx -19.7$ when  the critical value of $M/R$ for $Q/R = 1$ is zero. The quasi-periodic behavior for large negative $j$ that was found for the $b = 0$ models is also repeated and is shown in Figure \ref{fig:f23}. In the plots in  Figure \ref{fig:f23} we varied     $l = \sqrt{|6/5 + j|}$.

\subsection {Solutions with  b = 6. }
When $b =6$,  
  \begin{equation}
e^{-2 \lambda(r)} = 1-  \frac{2M}{R}  \frac{r^2}{R^2}  +  \frac{Q^2}{R^2} \frac{r^4}{R^4},
\end{equation}
and
 \begin{eqnarray}
  u(r) 
=  \frac{R}{2 Q} \left [ \log \left(\frac{M}{Q} + 1\right)  - \log \left( \frac{M}{Q} -   \frac{Q}{R^3} r^2 + e^{-\lambda(r)} \right) \right]
\end{eqnarray}  
and the master equation becomes
\begin{equation}
\label{zzz}
\frac{ d^2 \tilde{\zeta}}{d u^2}  = (1 + j) \frac{Q^2}{R^2}  \tilde{\zeta} (u),
\end{equation}
 \noindent The  3 distinct cases to be considered here  are  $j > -1, j =  -1 $ and $ j < -1$.

 \subsubsection{ Solutions for b = 6 and j > -1}
 \hspace{0.2 in}

\begin{figure}
\centering
  \includegraphics[width=0.6\linewidth]{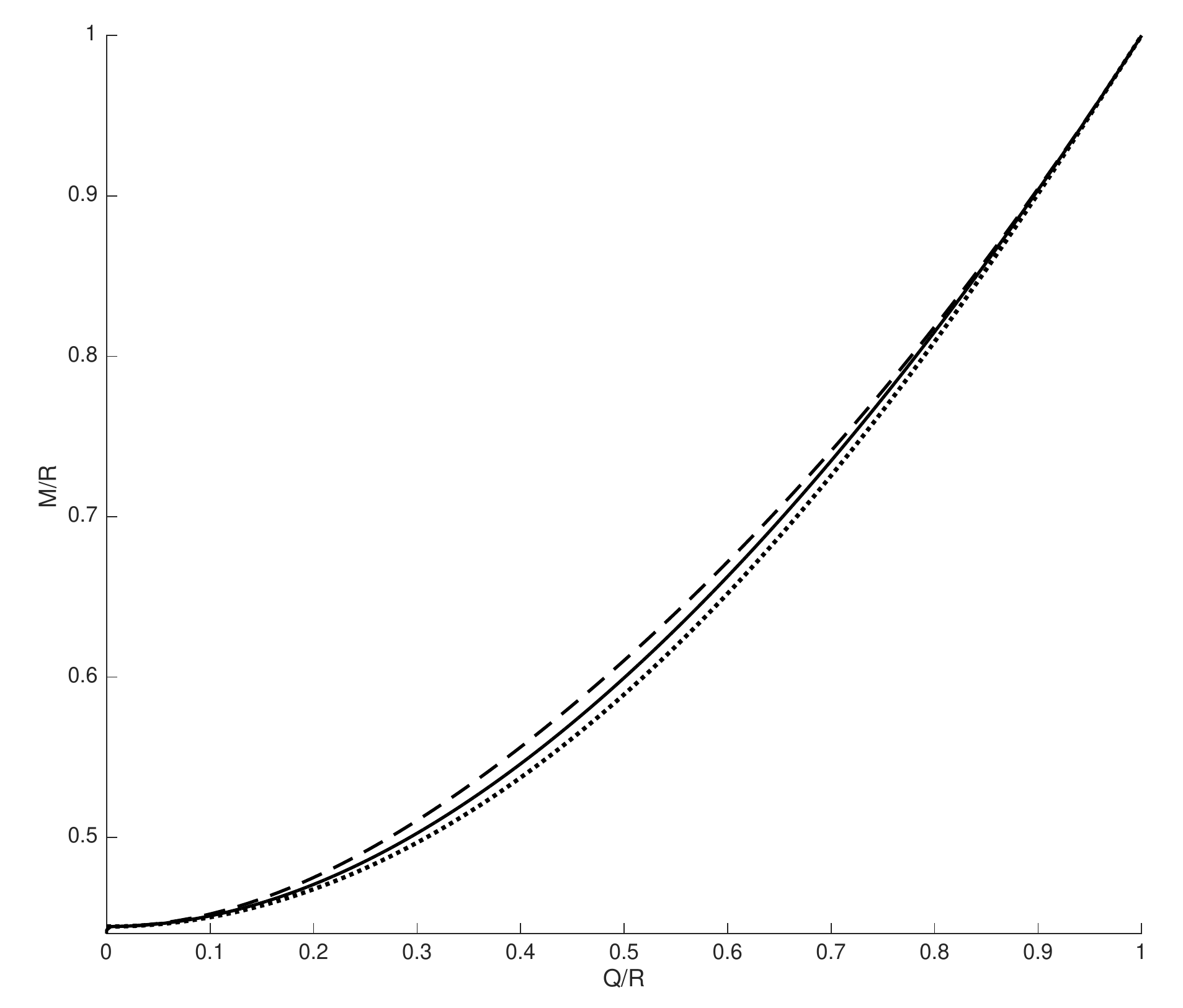}
  \caption{    The critical values of M/R vs Q/R from (\ref{aneq7}) for  $j = 8$   ({\bf - - - - -}), the  Andr\'{e}asson formula   (\rule{0.9 cm}{0.03 cm})  and  the isotropic model  ({\bf $\cdot$$\cdot$$\cdot$$\cdot$$\cdot$$\cdot$$\cdot$}) . 
        }
  \label{fig:f7}
\end{figure}

\noindent The solution for (\ref{zzz}) with $j  > - 1$   is 
\begin{equation}
\fl ~~~~~~~~~~~\tilde{\zeta} (u(r))  = A  \exp \left(p \frac{Q}{R} u(r)  \right)          +  B \exp \left(p \frac{Q}{R} u(r)  \right) ~~{\rm with} ~~p = (1 +j)^{\frac{1}{2}}
 \end{equation}
 with
 \begin{equation}
A = \frac{1}{2} \left[ \left(1 - \frac{2M}{R} + \frac{Q^2}{R^2}\right)^{\frac{1}{2}}  + \frac{1}{p} \left(\frac{M}{Q} - \frac{Q}{R}\right) \right] \exp \left (-p \frac{Q}{R} u(R)  \right),
\end{equation}
\begin{equation}
B = \frac{1}{2} \left[ \left(1 - \frac{2M}{R} + \frac{Q^2}{R^2} \right)^{\frac{1}{2}}  - \frac{1}{p} \left(\frac{M}{Q} - \frac{Q}{R}\right) \right] \exp\left( p \frac{Q}{R} u(R)  \right)
\end{equation}
 and
  \begin{eqnarray}
  u(R) 
=  \frac{R}{2 Q} \left [ \log \left(\frac{M}{Q} + 1\right)  - \log \left( \frac{M}{Q} -   \frac{Q}{R}  + e^{-\lambda(R)} \right) \right]
\end{eqnarray} 
 The critical values of $M/R$ are solutions of the following equation:
\begin{eqnarray}
\label{aneq7}
\fl ~~~~~~~~~\frac{ \left[          \left(\frac{M}{Q} - \frac{Q}{R}\right)           +      p \left(1 - \frac{2M}{R} + \frac{Q^2}{R^2}\right)^{\frac{1}{2}}               \right]} {
{ \left[        \left(\frac{M}{Q} - \frac{Q}{R}\right)           -      p \left(1 - \frac{2M}{R} + \frac{Q^2}{R^2}\right)^{\frac{1}{2}}           \right] }} =  \left[ \frac{(\frac{M}{Q} + 1)}{ \frac{M}{Q} - \frac{Q}{R} + \left(1 + \frac{2M}{R} + \frac{Q^2}{R^2}\right)^{\frac{1}{2}}} \right]^p
\end{eqnarray}
This critical values equations can be written as a polynomials and thus we were able to obtain values of $M/R$ for all $0 < Q/R < 1$. The critical values equation equation here has an exact solution for $p = 2$ i.e., $j = 3$. The expression is 
\begin{equation}
\label{aneqx}
\frac{M}{R} = \frac{1}{2}  + \frac{Q^2}{2R^2}  -  \frac{1}{2} \left[ \frac{\tilde{Q}}{9}   - \frac{9}{\tilde{Q}} \left(\frac{Q^2}{3R^2} - \frac{16}{81}\right) -\frac{5}{9}\right]^2
\end{equation}
 with
 \begin{equation}
 \tilde{Q} = \left[  64 + 81\frac{Q^2}{R^2}   + 9 \left(243 \frac{Q^6}{R^6}  - 351 \frac{Q^4}{R^4} + 384 \frac{Q^2}{R^2} \right)^{\frac{1}{2} }\right]^{\frac{1}{3}}
 \end{equation}
 The critical values of $M/R$ from this equation are slightly less that the corresponding values from  the  Andr\'{e}asson formula. 
 The results from the numerical solutions for (\ref{aneq7}) indicates that the $j =4$ model critical values saturates  the  Andr\'{e}asson formula. The $j =8$  curve is shown  in   Figure \ref{fig:f7}   and we see that the critical values  for this model are greater than the critical values from  the  Andr\'{e}asson formula.
 
 \subsubsection{ Solutions for b = 6 and j = -1.}
\hspace{0.2 in}

\begin{figure}
\centering
  \includegraphics[width=0.6\linewidth]{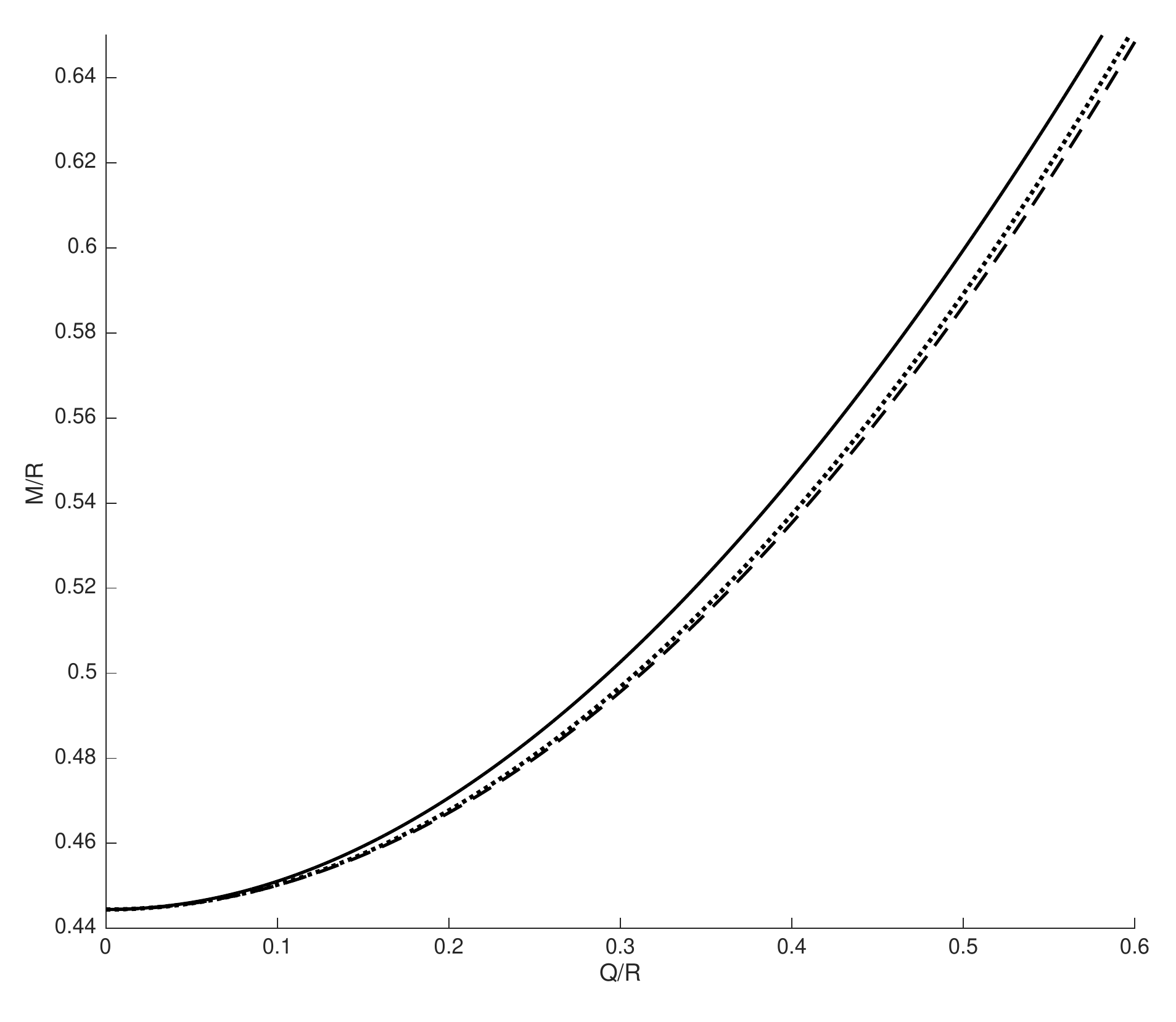}
  \caption{    The critical values of M/R vs Q/R for the $b = 6$ and  $j = -1$ model    ({\bf - - - - -}), the  Andr\'{e}asson formula   (\rule{0.9 cm}{0.03 cm})  and  the isotropic model  ({\bf $\cdot$$\cdot$$\cdot$$\cdot$$\cdot$$\cdot$$\cdot$}). 
        }
  \label{fig:f8}
\end{figure}

\noindent When  $b = 6$ and $j = -1$ there is only one solution for the master equation. Here
\begin{equation}
\frac{ d^2 \tilde{\zeta}}{d u^2}  = 0.
\end{equation} 
\noindent The solution for $\zeta(r)$ is 
\begin{eqnarray}
\fl ~~~~~~~~~~~~~~  \zeta(r) 
=\left(1 - \frac{2M}{R} + \frac{Q^2}{R^2}\right)^{\frac{1}{2}} + \frac{1}{2 } \left( \frac{M}{Q} - \frac{Q}{R}\right)\log  \left [  \frac{ 
  \frac{M}{Q} -   \frac{Q}{R}  + e^{\lambda(R)} }{     \frac{M}{Q} -   \frac{Q}{R^3} r^2 + e^{\lambda(r)} }                                  \right]
\end{eqnarray} 
Here the critical values of $M/R$ are solutions of the the following equation:
\begin{equation}
\label{aneq11}
\fl ~~~~~~~~~~~~\left( 1 - \frac{2M}{R}  + \frac{Q^2}{R^2} \right)^{\frac{1}{2}}
 = \frac{1}{2} \left( \frac{M}{Q} - \frac{Q}{R} \right) \log 
 \left[ \frac { \frac{M}{Q} + 1 }{  \frac{M}{Q} - \frac{Q}{R} + \left( 1 - \frac{2M}{R}  + \frac{Q^2}{R^2} \right)^{\frac{1}{2}}} \right]
\end{equation}
This equation was solved numerically and the results are plotted in Figure \ref{fig:f8}. We found that the critical values from  this model is less than values from  both the   the  Andr\'{e}asson formula  and the $b = 1$ isotropic model. 

 \subsubsection{ Solutions for b = 6 and  j < -1.}

\hspace{0.2 in} 
When  $b = 6$ and $j = -1$ the master equation becomes 
\begin{equation}
\frac{ d^2 \tilde{\zeta}}{d u^2}  = - t^2 \frac{Q^2}{R^2}  \tilde{\zeta} (u), ~~~{\rm with} ~~t = \sqrt{|1+j|}
\end{equation} 

The solution for $\tilde{\zeta}(u(r)$ here is 
\begin{equation}
\tilde{\zeta}(u(r)) = A \sin \left(t \frac{Q}{R} u(r) \right)  + B \cos \left(t  \frac{Q}{R} u(r) \right) 
\end{equation}
with 
 \begin{equation}
 A =     e^{-\lambda(R)}        \sin\left(t\frac{Q}{R} u(R) \right )         +  \frac{1}{t} \left(\frac{M}{Q} - \frac{Q}{R} \right) \cos\left(t\frac{Q}{R} u(R) \right )  ,
\end{equation}

\begin{equation}
  B =     e^{-\lambda(R)}     \cos\left(t\frac{Q}{R} u(R) \right )        - \frac{1}{t} \left(\frac{M}{Q} - \frac{Q}{R} \right) \sin\left(t\frac{Q}{R} u(R) \right ).
\end{equation}

\begin{figure}
\centering
 \includegraphics[width=0.6\linewidth]{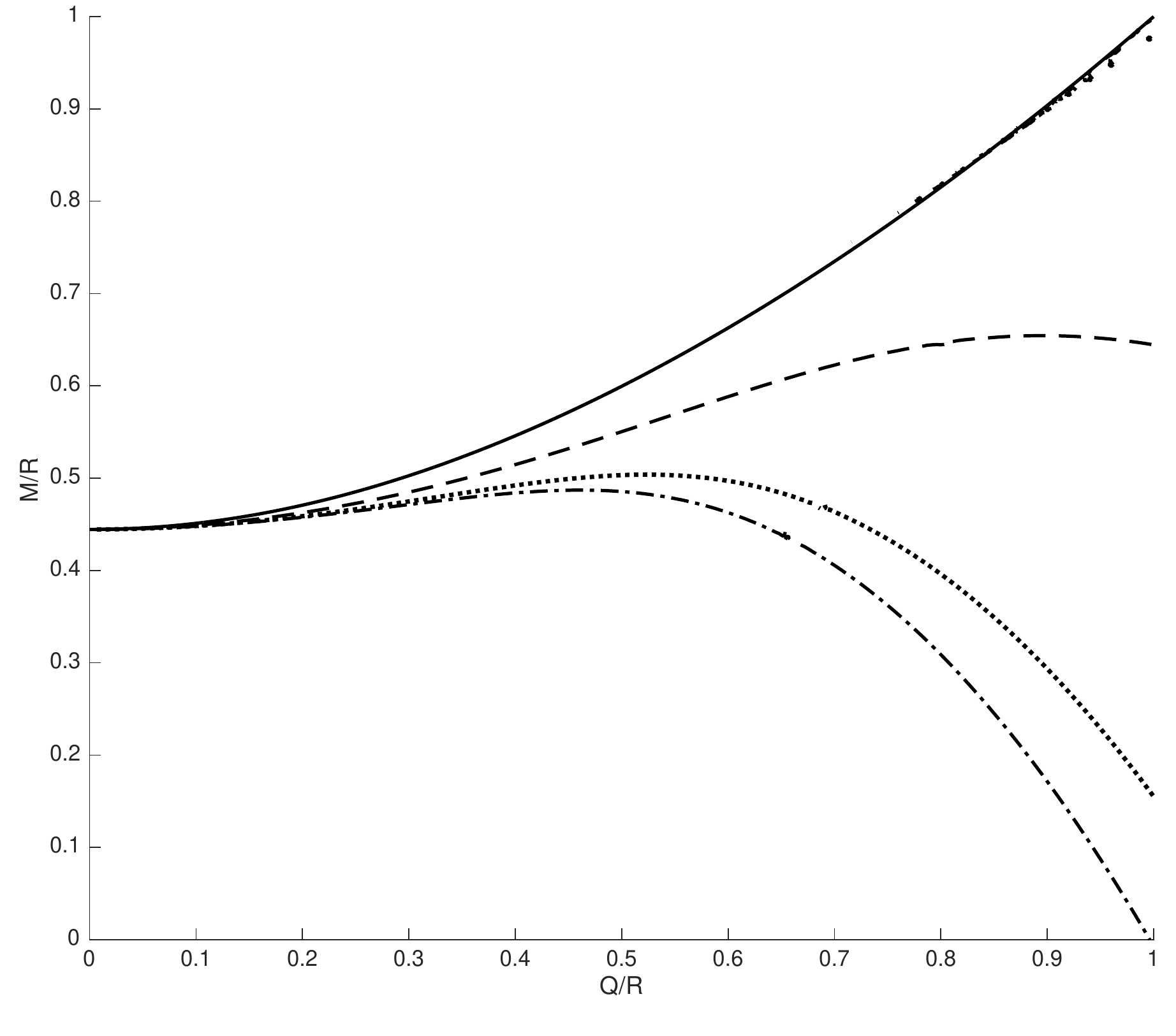}
  \caption{  The critical values of M/R vs Q/R  for  $j = -9$      $({\bf- - - -})$,     $j = -15$   $({\bf \cdots \bf \cdots        }) $               and    $j = -16.8$    $({\bf \cdot-\cdot-\cdot})$     from (\ref{aneq101})    and     the  Andr\'{e}asson formula   (\rule{0.9 cm}{0.03 cm}). }
  \label{fig:f9}
\end{figure}
\begin{figure}[h!]
  \centering
  \begin{subfigure}[b]{0.45\linewidth}
    \includegraphics[width=\linewidth]{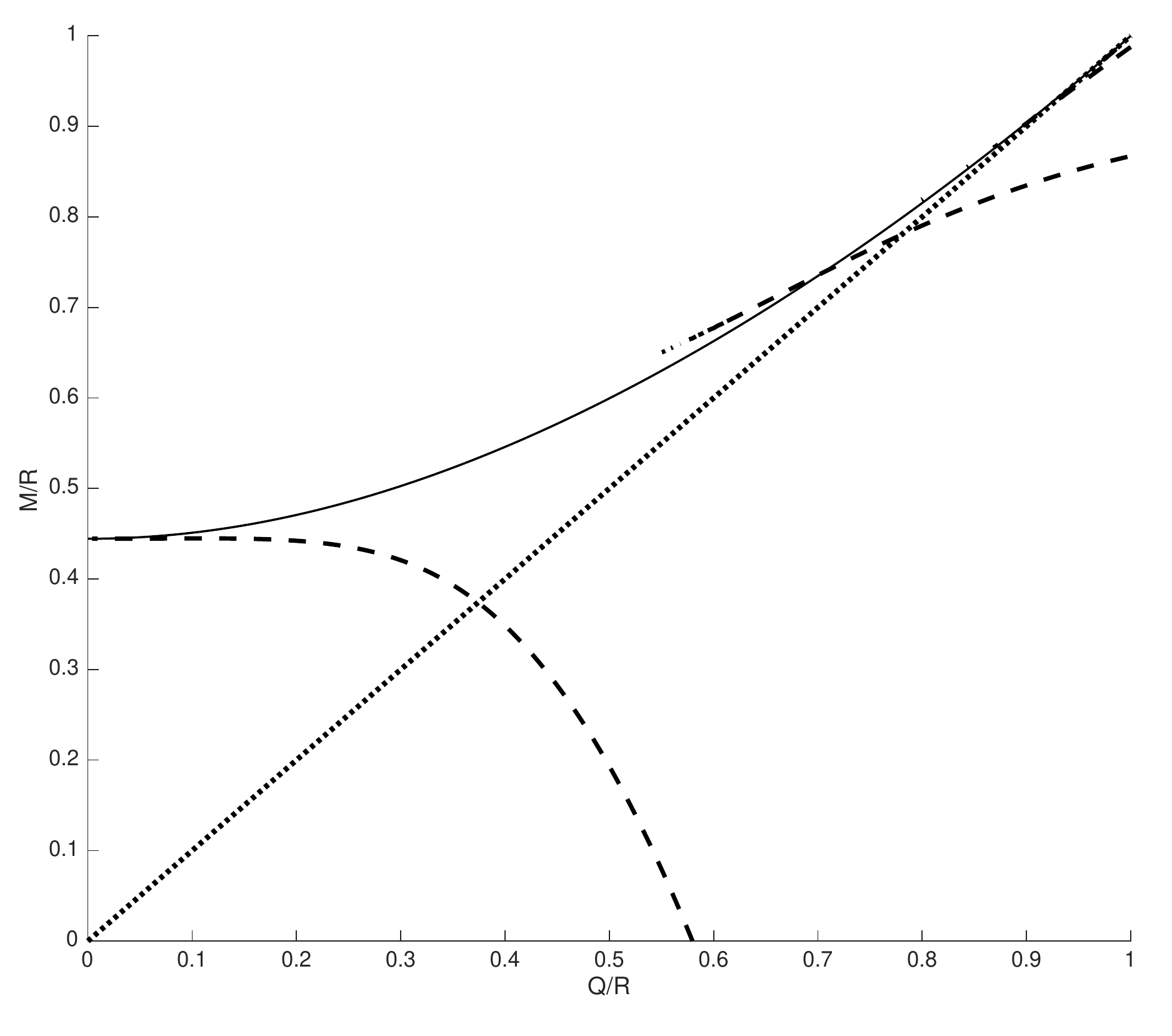}
     \caption{ $t = 6$.}
  \end{subfigure}
  \begin{subfigure}[b]{0.45\linewidth}
    \includegraphics[width=\linewidth]{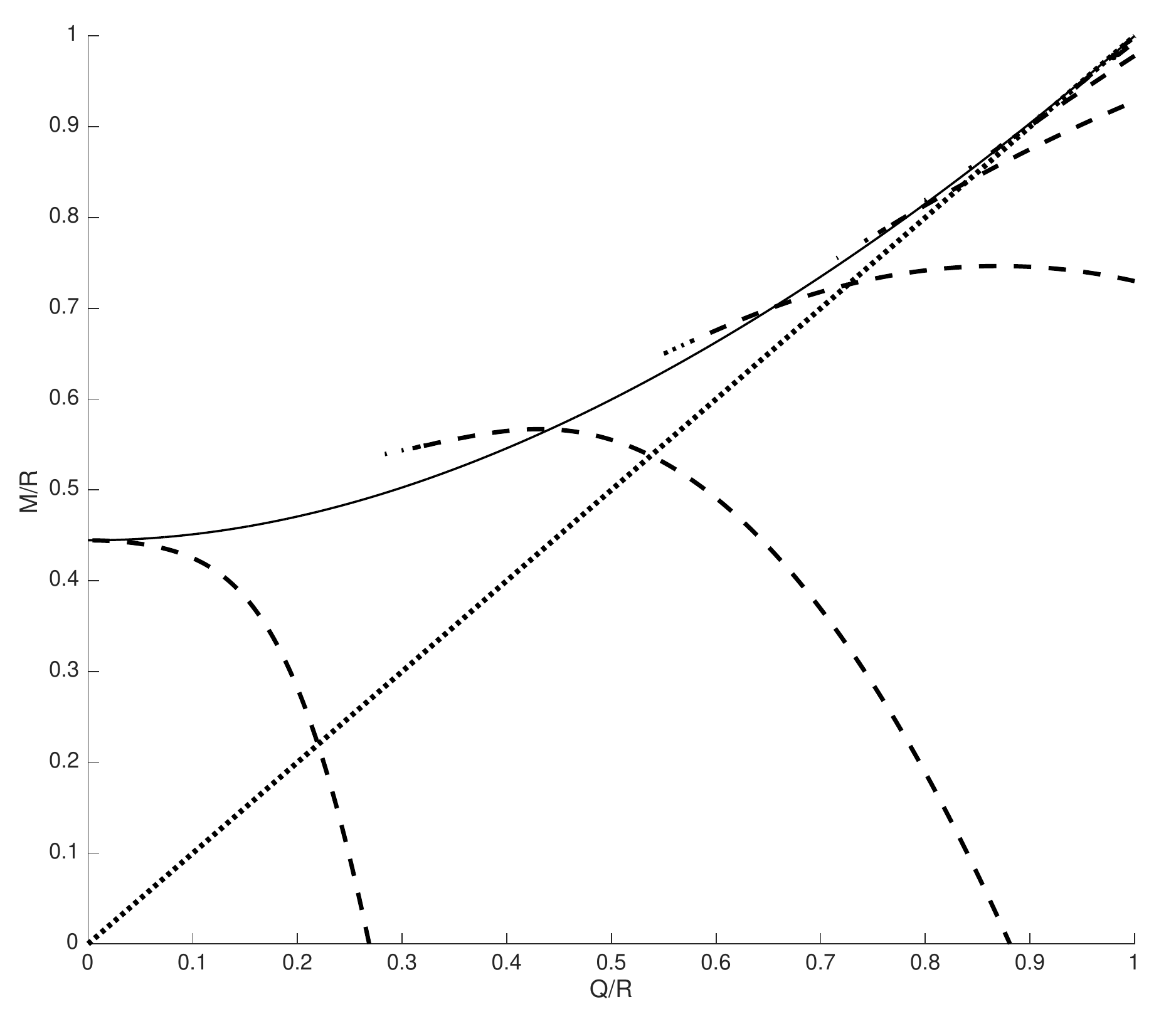}
    \caption{$t = 12.$}
  \end{subfigure}
  \begin{subfigure}[b]{0.45\linewidth}
    \includegraphics[width=\linewidth]{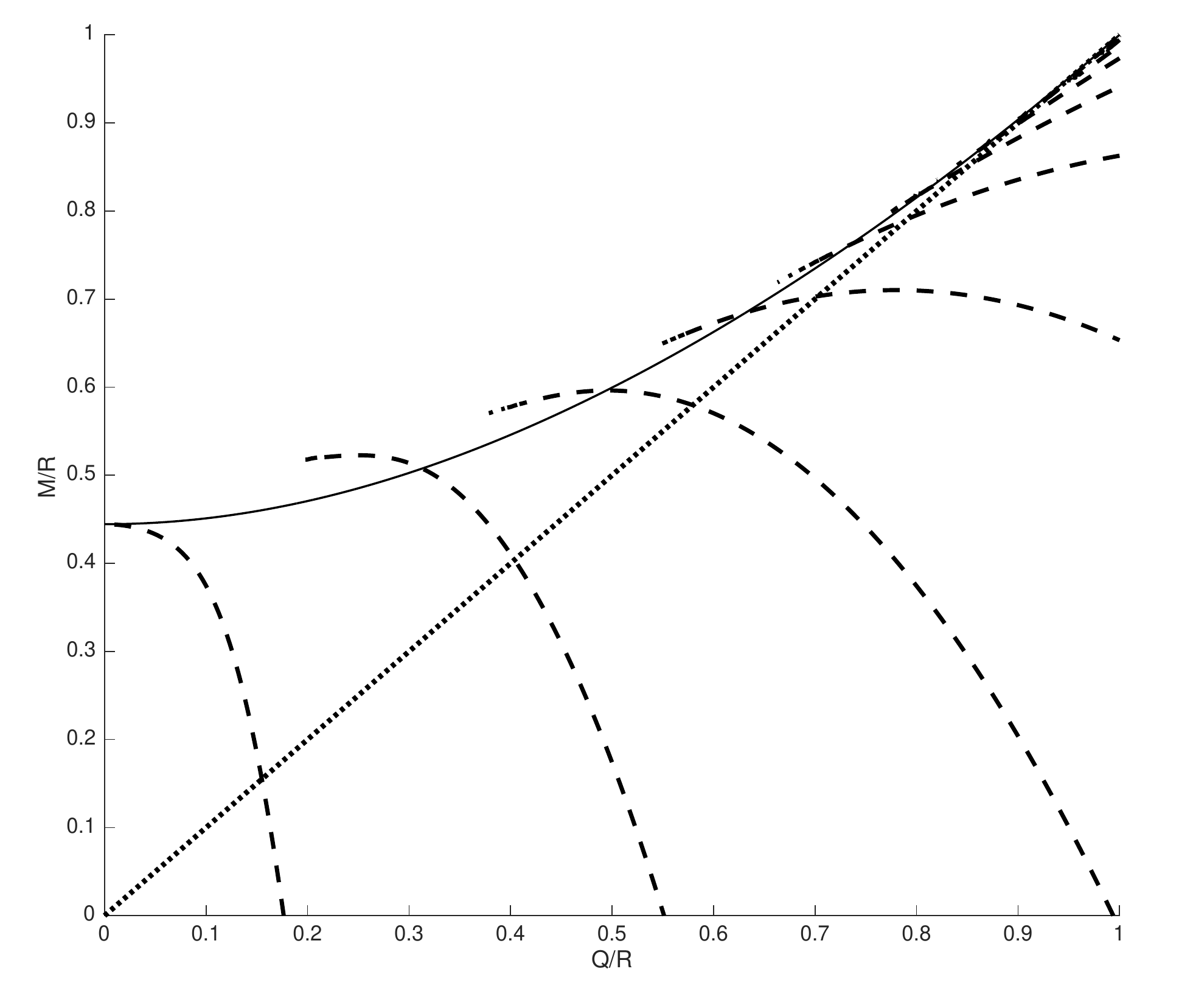}
    \caption{$t = 18.$}
  \end{subfigure}
  \begin{subfigure}[b]{0.45\linewidth}
    \includegraphics[width=\linewidth]{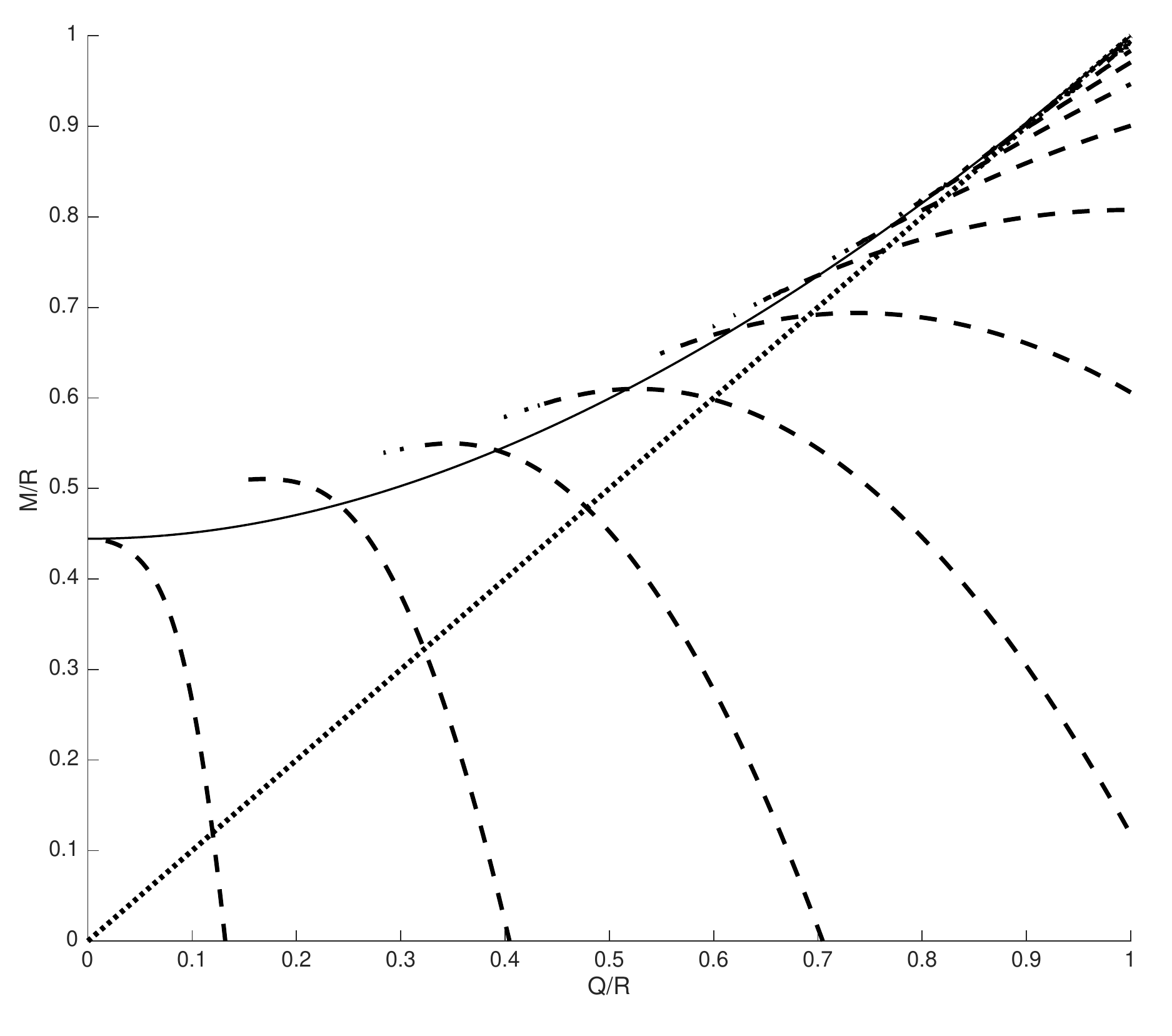}
    \caption{$ t = 24$.}
  \end{subfigure}
    \caption{ \it The critical values curves (${\bf ----}$) that are obtained from (\ref{aneq101})  by varying $t$. The dotted line   $({\bf \cdot \cdot \cdot \cdot \cdot})$     is the  $M/R = Q/R$ extremal line     and     the curve from  Andr\'{e}asson formula is the solid line  (\rule{0.9 cm}{0.03 cm}). } 
  \label{fig:f27}
\end{figure}

\noindent The stability condition $\zeta(r= 0) = 0$ requires
\begin{equation}
 A  \sin\left(t\frac{Q}{R} u(0) \right )           +  B \cos \left( t\frac{Q}{R} u(0)  \right) = 0,
 \end{equation} 
however, since $u(0) \equiv 0$, then  $B$ must be equal to zero here. This results in   the following critical values equation:
\begin{equation}
 e^{-\lambda(R)}     \cos\left(t\frac{Q}{R} u(R) \right )        = \frac{1}{t} \left(\frac{M}{Q} - \frac{Q}{R} \right) \sin\left(t\frac{Q}{R} u(R) \right ) 
 \end{equation}
Since for $b = 6$ 
\begin{eqnarray}
u(R)    
=  \frac{R}{2 Q} \left [ \log \left(\frac{M}{Q} + 1\right)  - \log \left( \frac{M}{Q} -   \frac{Q}{R}  + e^{-\lambda(R)} \right) \right],
 \end{eqnarray}
 The critical values of $M/R$ here are solutions of the following equation:
 \begin{equation}
\label{aneq101}
 \fl ~~ \left( 1 - \frac{2M}{R}  + \frac{Q^2}{R^2} \right)^{\frac{1}{2}}     = \frac{1}{t} \left(\frac{M}{Q} - \frac{Q}{R} \right) \tan\left(\frac{t}{2}  \log \left[ \frac{\left(\frac{M}{Q} + 1\right) } { \left( \frac{M}{Q} -   \frac{Q}{R}  +\left( 1 - \frac{2M}{R}  + \frac{Q^2}{R^2} \right)^{\frac{1}{2}} \right)} \right]\right ) 
 \end{equation}
The numerical solution of this equation is plotted in Figure \ref{fig:f9} for various values of $j$. The behavior of the critical curves here are similar to the $b = 0$ and $b = 1$ models. However, here the critical values of $M/R$  for a given $Q/R$ and $j$ are smaller than either the $b = 0$ or the $b = 1$ models. In particular here when $j = -16.8$ the critical value of $M/R$ for $Q/R = 1$ is zero.
The quasi-periodic behavior for large negative $j$ or $t = \sqrt{|1/5 + j|}$ is shown in Figure \ref{fig:f27}.

 \subsection{Solutions with 11/5 -  b/5  + j   = 0.}  
 \hspace{0.2in}
 We now consider a class of solutions that satisfies the condition $11/5 - b/5 + j = 0$. The special cases $b = 0$, $b = 1$ and $b = 6$ were already studied above. Here and below we will use 
 \begin{equation}
 \ba \equiv \frac{b}{5} - \frac{ 1}{5},
 \end{equation}
  thus $11/5 - b/5 +j$ becomes $2 -\ba +j$. We will start with the case  $\ba > 0$.
 
\subsubsection{  Solutions with (2 - $\ba$  + j)     = 0  and $ \ba$ > 0}
\hspace{0.2in}
The following general expressions exists for  $\ba  > 0$: 
\begin{equation}
\label{lambda1}
e^{-2 \lambda(r)} = 1-  \left( \frac{2M}{R}  + (\ba-1) \frac{Q^2}{R^2}    \right) \frac{r^2}{R^2}  +  \ba \frac{Q^2}{R^2} \frac{r^4}{R^4},
\end{equation}
\begin{eqnarray}
\label{urx}
 \fl ~~~~~~~u(r)  = \frac{1}{R^2} \int_0^r s e^{\lambda(s) } ds   = \frac{1}{2 \sqrt{\ba}}\frac{R}{Q}\left[  \log \left( \frac{1}{\sqrt{\ba}}\frac{M}{Q} + \frac{(\ba -1)}{2 \sqrt{\ba}} \frac{Q}{R} + 1\right) \right] \\ \nonumber
~~~~~~~~~~~~~~~ -   \frac{1}{2 \sqrt{\ba}}\frac{R}{Q}\left[     \log \left( \frac{1}{\sqrt{\ba}}\frac{M}{Q} + \frac{(\ba -1)}{2 \sqrt{\ba}} \frac{Q}{R} -\frac{1}{\sqrt{\ba}} \frac{Q r^2}{R^3}   + e^{-\lambda(r)} \right)\right]
\end{eqnarray}
 and
\begin{eqnarray}
\label{ur1}
\fl {\zeta} (u(r))   = A  +  B  u(r) = \frac{1}{2 \sqrt{\ba}} \left(\frac{M}{Q} - \frac{Q}{R} \right)
 \left[     \log \left( \frac{1}{\sqrt{\ba}}\frac{M}{Q} + \frac{(\ba -3)}{2 \sqrt{\ba}} \frac{Q}{R}    + e^{-\lambda(R)} \right)\right.  \\ \nonumber
\fl -                        \left.     \log \left( \frac{1}{\sqrt{\ba}}\frac{M}{Q} + \frac{(\ba -1)}{2 \sqrt{\ba}} \frac{Q}{R} -\frac{1}{\sqrt{\ba}} \frac{Q r^2}{R^3}   + e^{-\lambda(r)} \right)\right]   + 
\left (1  -  \frac{2M}{R}   + \frac{Q^2}{R^2}\right)^{\frac{1}{2}}  
\end{eqnarray}
The condition $\zeta(r = 0) = 0$ gives the following transcendental for the critical values of $M/R$ as a function of $Q/R$:
\begin{eqnarray}
\label{aneq15}
\fl \left (1  -  \frac{2M}{R}   + \frac{Q^2}{R^2}\right)^{\frac{1}{2}}           = \frac{1}{2 \sqrt{\ba}} \left(\frac{M}{Q} - \frac{Q}{R} \right)  \left[  \log \left(  \frac{1}{\sqrt{\ba}}\frac{M}{Q} + \frac{(\ba -1)}{2 \sqrt{\ba}} \frac{Q}{R}    + 1\right) \right. \\ \nonumber
  -  \left.  \log    \left(        \frac{1}{\sqrt{\ba}}\frac{M}{Q} + \frac{(\ba -3)}{2 \sqrt{\ba}} \frac{Q}{R}    + e^{-\lambda(R)}\right)  \right].  
 \end{eqnarray}
  
 \begin{figure}
\centering
 \includegraphics[width=0.6\linewidth]{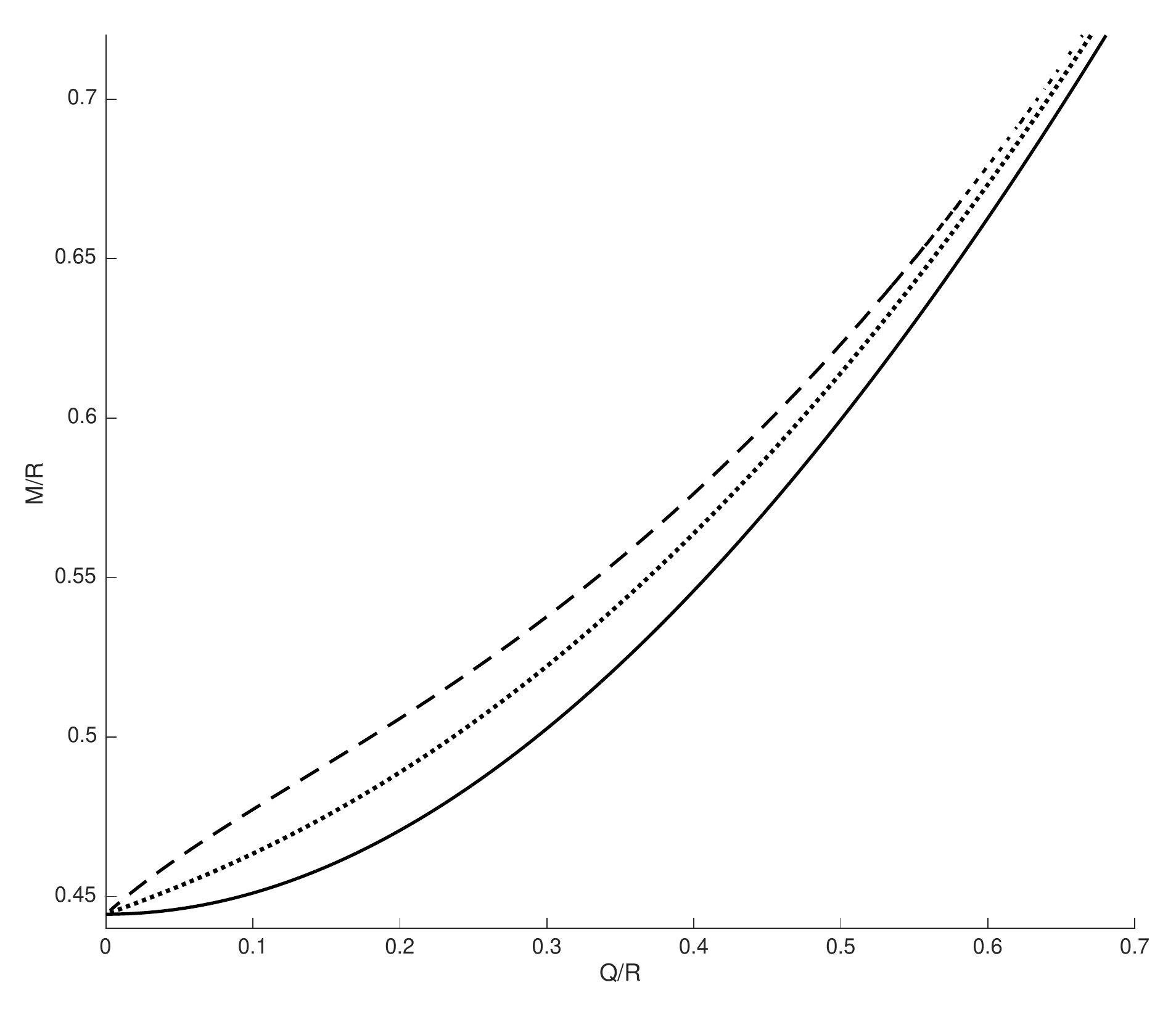}
 \caption{  The critical values of M/R vs Q/R  for   $\ba = 16$   $({\bf ---})$ and   $\ba = 4$   $({\bf \cdots \bf \cdots        }),$   from (\ref{aneq15})    and     the  Andr\'{e}asson formula   (\rule{0.9 cm}{0.04 cm}).      }
\label{fig:f10}
\end{figure}
 \begin{figure}
\centering
 \includegraphics[width=0.6\linewidth]{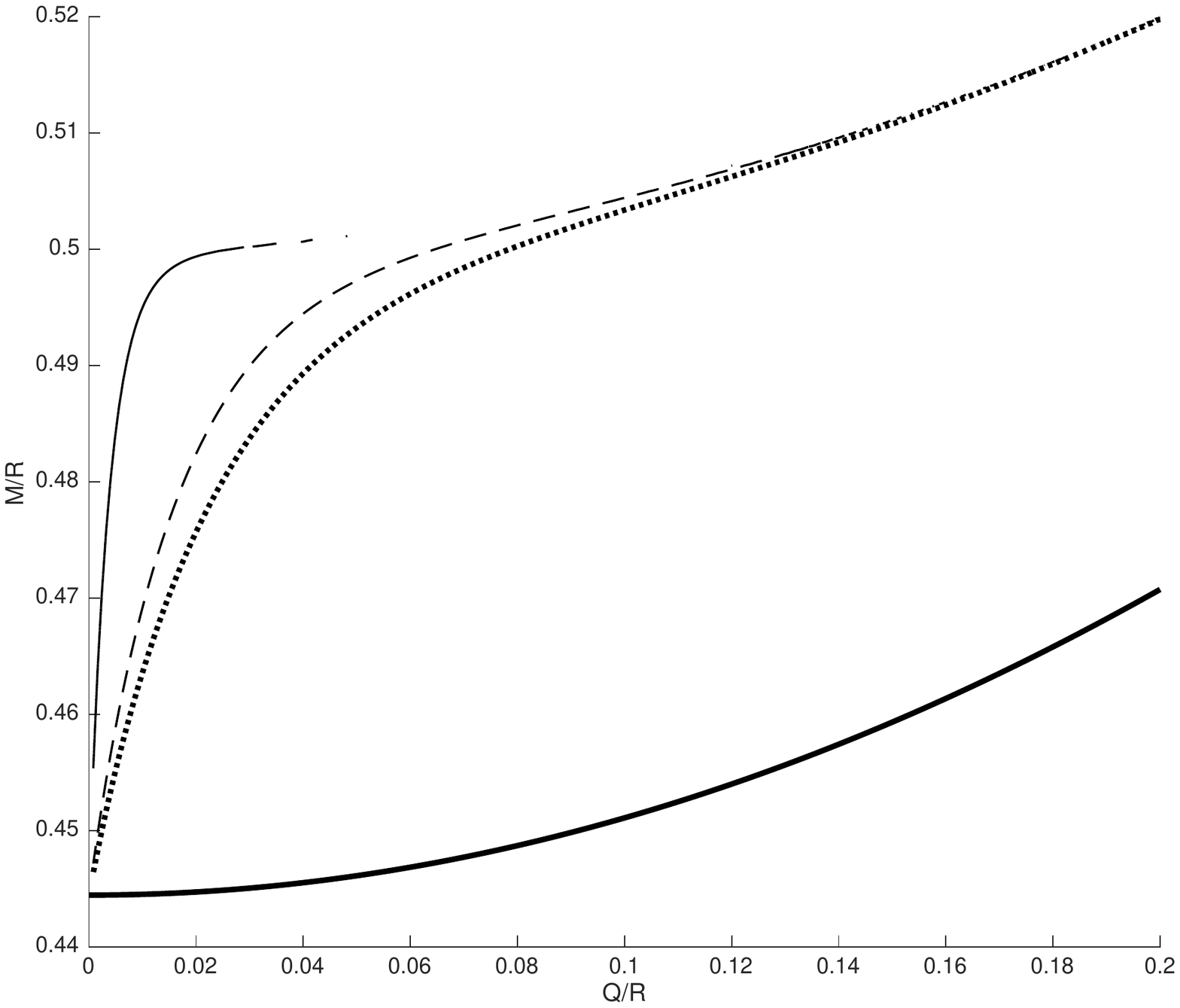}
 \caption{  The critical values of M/R vs Q/R  for      $\ba = 20000$  (\rule{0.9 cm}{0.01 cm}),   $\ba = 1000$   $({\bf ---})$,   and          $\ba = 500$ $({\bf \cdots \bf \cdots  }),$  from (\ref{aneq15})    and     the  Andr\'{e}asson formula   (\rule{0.9 cm}{0.04 cm}).      }
\label{fig:f11}
\end{figure}

\noindent  We solved this equation  numerically and the results are plotted in Figure \ref{fig:f10} and Figure \ref{fig:f11} . We found that when $0 < \ba < 24$ the curves (Figure \ref{fig:f10}) are similar to the  curve from the 
 Andr\'{e}asson's               formula. The curves with $\ba < 1.5$ have critical values of $M/R$ that are less than the corresponding values from the  Andr\'{e}asson's               formula.
 The critical values of  $M/R$  for $\ba = 1.5$ almost matches the values from the  Andr\'{e}asson's               formula and the curves for $\ba > 1.5$ have critical values of $M/R$ that are greater than the values from the   Andr\'{e}asson's               formula.  We note note however the range of $Q/R$ for which a solution exists get progressively smaller with increasing $\ba$.  For large    $\ba$  ($\ba > 500$)  in  Figure \ref{fig:f11}   we observe that  the critical values of $M/R$ for a given $\ba$ increase very rapidly with increasing values of $Q/R$ but as we have just noted the range of values of $Q/R$ for which a solution exists is small,  for example when $\ba =  20000$, $M/R$ values exists only for the $0 < Q/R < 0.04$.

\subsubsection{  Solutions with (2 -  $\ba$  + j)     = 0\,  and \,$\ba$ < 0.}
$~~~$
\vspace{0.1 in}

\noindent We will now develop solutions for \ref{a1eq} for with $\ba <0$. We can write the following general expressions for these solutions using $|\ba| = k$
\begin{equation}
e^{-2 \lambda(r)} = 1-  \left( \frac{2M}{R}  - (k +1)  \frac{Q^2}{R^2}    \right) \frac{r^2}{R^2}  -  k  \frac{Q^2}{R^2} \frac{r^4}{R^4},  ~~~~ k = |\ba|
\end{equation}
and
\begin{eqnarray}
\fl u(r) =    \frac{1}{R^2} \int_0^r   e^{\lambda(s)}  s ds                               = \frac{1}{R^2} \int_0^r \frac{s ds }{  \left[1-  \left( \frac{2M}{R}  - (k + 1)   \frac{Q^2}{R^2}    \right) \frac{s^2}{R^2}    -k  \frac{Q^2}{R^2} \frac{s^4}{R^4} \right]^{\frac{1}{2}}}\\
~~~~~~~~~~~\fl = \frac{1}{2 \sqrt{k}}\frac{R}{Q} \left[ \tan^{-1}  \left(  \left(            \frac{1}{\sqrt{k}}  \frac{M}{Q} -  \frac{k +1}{2\sqrt{k}}\frac{Q}{ R }+  \sqrt{k}  \frac{Q }{R^3} r^2 \right) e^{\lambda(r)} \right)  \right. \\ \nonumber
~~~~~~~~~~~~~~~~~~~~~~~~~~~~~~~~~~~~~~~~~~~~~~\left.  -\tan^{-1}       \left(  \frac{1}{\sqrt{k}}  \frac{M}{Q} -  \frac{k +1}{2\sqrt{k}}\frac{Q}{ R }\right)\right].
\end{eqnarray}
\begin{eqnarray}
\fl {\zeta} (u(r))   = \frac{1}{2 \sqrt{k}} \left(\frac{M}{Q} - \frac{Q}{R} \right) \left[ \tan^{-1}  \left(  \left(            \frac{1}{\sqrt{k}}  \frac{M}{Q} -  \frac{k +1}{2\sqrt{k}}\frac{Q}{ R }+  \sqrt{k}  \frac{Q }{R^3} r^2 \right) e^{\lambda(r)} 
\right)  \right. \\ \nonumber
\left.  -\tan^{-1}  \left(     \left(  \frac{1}{\sqrt{k}}  \frac{M}{Q} + \frac{k -1}{2\sqrt{k}}\frac{Q}{ R }\right) e^{\lambda(R)}\right)\right]  + \left(1 - \frac{2M}{R} + \frac{Q^2}{R^2}\right)^{\frac{1}{2}}
\end{eqnarray}
The stability condition $\zeta(r(=0) \, = \,0$ gives the following  implicit equation for the dependence of the critical values of $M/R$ on $Q/R$: 
\begin{eqnarray}
\label{aneq12}
\fl ~~~~~\left(1 - \frac{2M}{R} + \frac{Q^2}{R^2}\right)^{\frac{1}{2}}  = \frac{1}{2 \sqrt{k}} \left(\frac{M}{Q} - \frac{Q}{R} \right) \left[ 
 \left(  \tan^{-1}  \left(     \left(  \frac{1}{\sqrt{k}}  \frac{M}{Q} + \frac{k -1}{2\sqrt{k}}\frac{Q}{ R }\right) e^{\lambda(R)}\right) \right. \right.
  \\ \nonumber
~~~~~~~~~~~~~~~~~~~~~~~~~~~~~~~~~~~~~~~~~~\left.   -\tan^{-1}  \left(             \frac{1}{\sqrt{k}}  \frac{M}{Q} -  \frac{k +1}{2\sqrt{k}}\frac{Q}{ R } \right) \right]
 \end{eqnarray}
 
 \begin{figure}
\centering
 \includegraphics[width=0.6\linewidth]{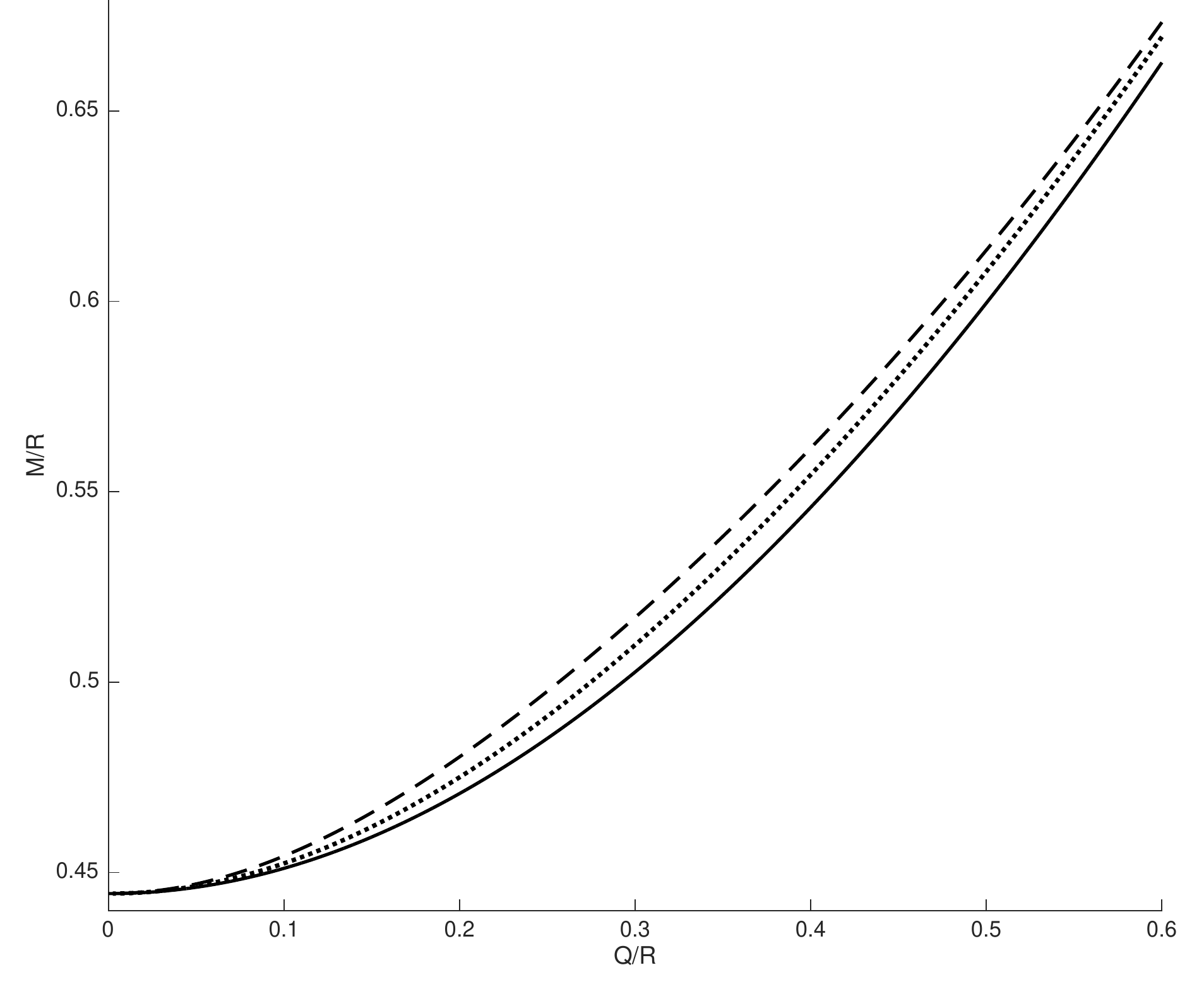}
 \caption{  The critical values of M/R vs Q/R  for   $|\ba| = 32$   $({\bf ---})$ and   $|\ba| = 16$   $({\bf \cdots \bf \cdots        }),$   from (\ref{aneq12})    and     the  Andr\'{e}asson formula   (\rule{0.9 cm}{0.04 cm}).      }
\label{fig:f12}
\end{figure}
 \begin{figure}
\centering
 \includegraphics[width=0.6\linewidth]{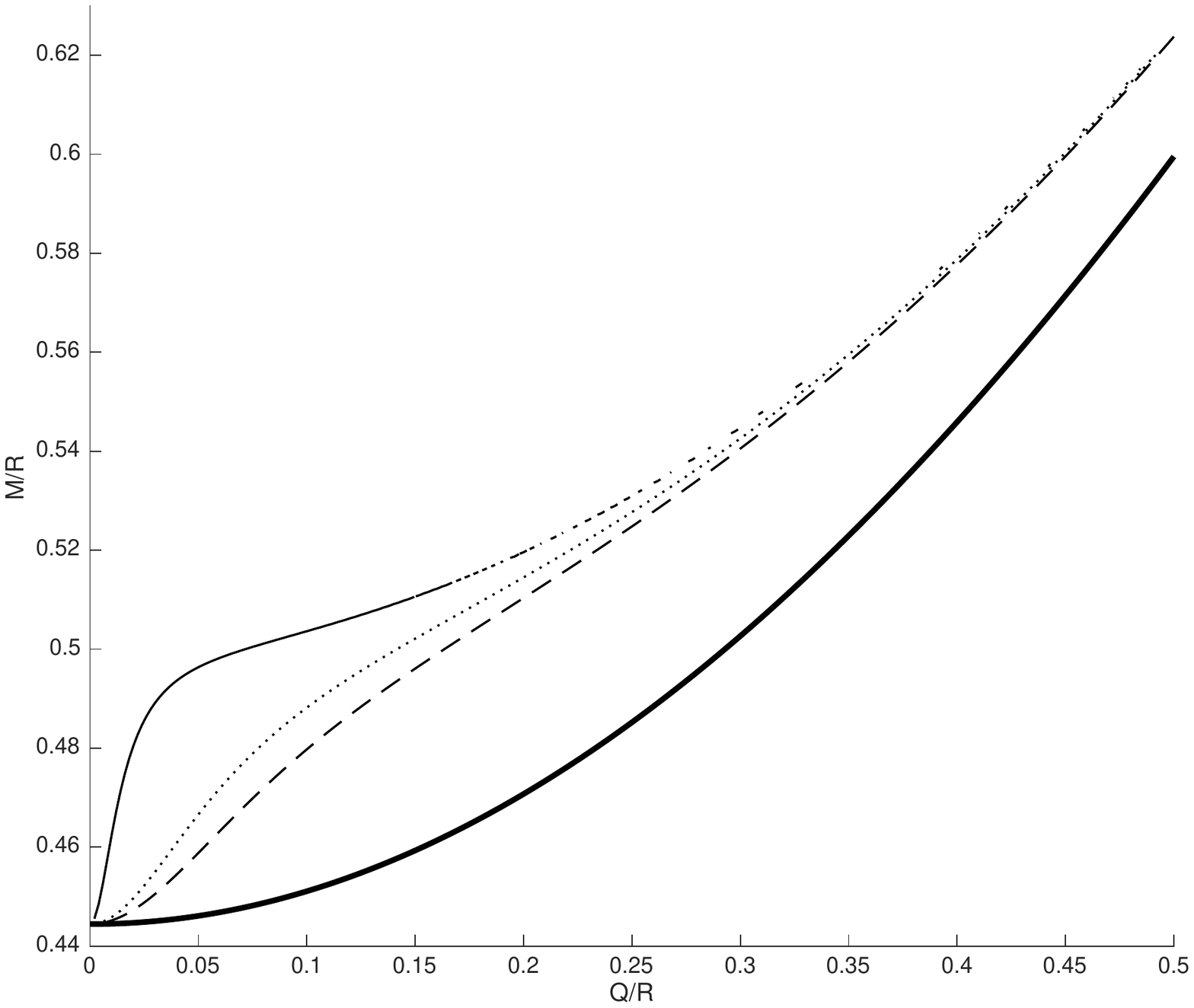}
 \caption{  The critical values of M/R vs Q/R  for      $|\ba| = 20000$  (\rule{0.9 cm}{0.01 cm}),   $|\ba| = 1000$   $({\bf \cdots \bf \cdots  }),$     and          $|\ba| = 500$ $(---)$ from (\ref{aneq15})    and     the  Andr\'{e}asson formula   (\rule{0.9 cm}{0.04 cm}).      }
\label{fig:f13}
\end{figure} 
 
We solved this equation numerically and the results are plotted in Figure \ref{fig:f12}.  We found that when $|\ba| \approx 8$  the critical values of $M/R$ approximately equal to corresponding values of $M/R$ from  the Andr\'{e}asson   formula and curves with $|\ba| > 8$ have critical $M/R$ values greater than the values from the  the Andr\'{e}asson   formula. We also found that the curves for  large $-\ba$ values have a similar pattern to the large $\ba$ values curves.  We note however that for large $-\ba$ values the range of $Q/R$  for which a solution for (\ref{aneq12}) exists is larger.

\subsection {Solutions with $( 2 - \ba + j)^{\frac{1}{2}}  \,     = \, a  \, and \, \ba = a^2/4$.  }
We studied  a set  of solutions with the condition  $( 2 - \ba + j) ^{\frac{1}{2}} \,     = \, a$  and $\ba = a^2/4.$ 
The general expressions for $e^{- 2 \lambda}$ and $u(r)$ here are same as (\ref{lambda1}) and (\ref{urx}). The solution for $\zeta(r) $ here is
\begin{equation*} 
 \tilde{\zeta} (u(r)) = A \exp\left({a \frac{Q}{R} u(r)} \right) +  B  \exp\left({-a \frac{Q}{R} u(r)} \right)  
\end{equation*}
The constants $A$ and $B$ are evaluated using the boundary conditions. Applying the boundary conditions we find that
\begin{equation}
A = \frac{1}{2} \left[ \left(1 - \frac{2M}{R} + \frac{Q^2}{R^2} \right)^{\frac{1}{2} } + \frac{1}{a} \left(\frac{M}{Q} - Q \right) \right] \exp\left({-a \frac{Q}{R} u(r)} \right) 
 \end{equation}
and
\begin{equation}
B = \frac{1}{2} \left[ \left(1 - \frac{2M}{R} + \frac{Q^2}{R^2} \right)^{\frac{1}{2} } - \frac{1}{a} \left(\frac{M}{Q} - Q \right) \right]  \exp\left({a \frac{Q}{R} u(r)} \right)  
\end{equation}
The stability condition  $\tilde{\zeta}(r \, = \, 0) \, = \, 0$ requires $A = -B$. Substituting $\ba = a^2/4$ and the expression for $u(R)$  into the equation $A = -B$ results in  the following equation for the critical values of $M/R$ as a function of $Q/R$ :

\begin{equation}
\label{aneq21}
\fl ~~~~~~~~~~\frac{\left[  \left(\frac{M}{Q} - Q \right)      + 2 \sqrt{\ba}  \left(1 - \frac{2M}{R} + \frac{Q^2}{R^2} \right)^{\frac{1}{2} }   \right]} {\left[\left(\frac{M}{Q} - Q \right)   - 2\sqrt{\ba}\left(1 - \frac{2M}{R} + \frac{Q^2}{R^2} \right)^{\frac{1}{2} }                   \right]}  =    \left[ \frac{  \frac{M}{Q} + \frac{(\ba -1)}{2 } \frac{Q}{R}    + \sqrt{\ba}} 
{ \frac{M}{Q} + \frac{(\ba -3)}{2 } \frac{Q}{R}    +\sqrt{\ba} e^{-\lambda(R)} } \right]^2.  
\end{equation}
The $\ba = 1$  equation here is the same as the $b = 6$ with $j = 3$ model, ({\ref{aneq7}) and thus has the exact solution (\ref{aneqx}).


\noindent The expressions from which the critical values of $M/R$ are found  for other values of $\ba$  can be written as polynomials, for example when  $\ba = 2$  the expression is 
\begin{eqnarray}
\fl ~~~~~~~~~\left( - 11\sqrt{2} \frac{M^2}{R^2}\frac{Q}{R} + 12\sqrt{2} \frac{M}{R}\frac{Q^3}{R^3}  + 4\sqrt{2} \frac{M}{R}\frac{Q}{R}        -3\sqrt{2}\frac{Q^5}{R^5} 
     \right. \\ \nonumber \left. 
     \fl ~~~~~~~~~~~~~~~~~~~~~~~~~~~~~~~~~~~~~~~~~~~~~~~~~~~~~ -2\sqrt{2} \frac{Q^3}{R^3}   
   -  2\frac{M^2}{R^2}     + \frac{M}{R} \frac{Q^2}{R^2} 
   + \frac{Q^4}{R^4}         \right)^2 =  \\ \nonumber
   \fl ~~~~~~~~~\left( - 6\frac{M^2}{R^2} + 11 \frac{M}{R} \frac{Q^2}{R^2} 
    - 4\sqrt{2} \frac{M}{R} \frac{Q}{R} - 6\frac{Q^4}{R^4}   - 2\frac{Q^3}{R^3} -  8\frac{Q^2}{R^2} \right)^{2} \left(1 - 2\frac{ M}{R} +\frac{Q^2}{R^2}\right)
               \end{eqnarray}

\begin{figure}
\centering
 \includegraphics[width=0.6\linewidth]{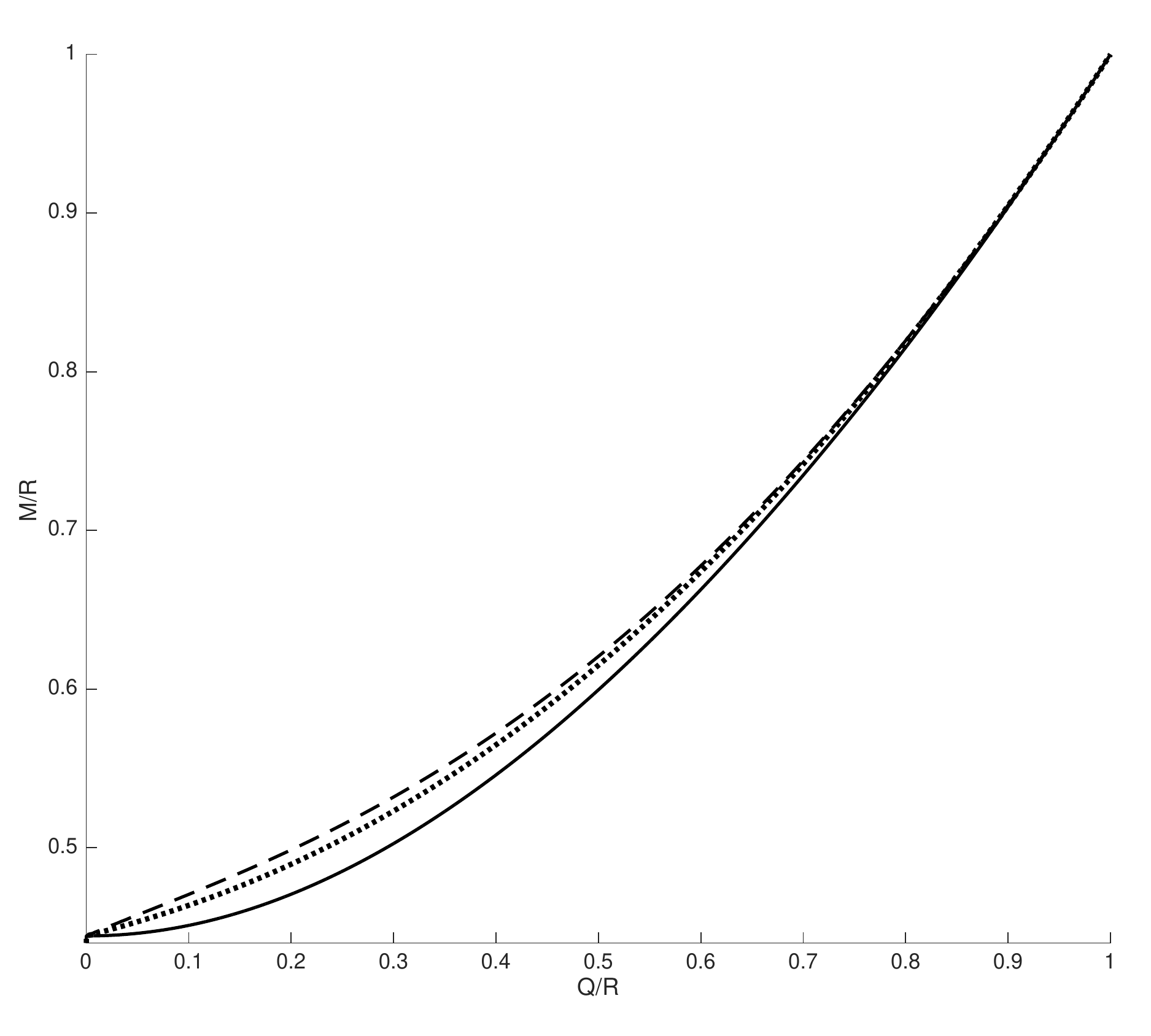}
 \caption{  The critical values of M/R vs Q/R  for   $\ba = 6$   $({\bf ---})$ and   $\ba = 4$   $({\bf \cdots \bf \cdots        }),$   from (\ref{aneq21})    and     the  Andr\'{e}asson formula   (\rule{0.9 cm}{0.04 cm}).      }
\label{fig:f14}
\end{figure}
 
\noindent   The numerical solution of  (\ref{aneq21}) shows that  for $\ba \, =\,1$ the $M/R$ values are less than those from           the     Andr\'{e}asson           formula but the for $\ba \, \geq 2 $ the $M/R$ values from from this model are greater than those 
from the        Andr\'{e}asson           formula for a given $Q/R$. We plot the curves for $\ba = 4$ and $\ba = 6$ in  Figure \ref{fig:f14}.


\section{Conclusion}

In this paper we studied a plethora of exact solutions for anisotropic charged spheres with the fluid density, the charge density and  the anisotropy given by the following functions respectively: 
\begin{equation}
\fl ~~~~~~~~~~~~\rho(r) = \rho_o - \frac{b  }{8 \pi}  \frac{Q^2}{R^6} r^2,  {\rm ~~~}  q(r) =   \frac{Q}{R^3} r^3, {\rm~~~and~~~} 8 \pi (p_t - p_r) = j   \frac{Q^2}{R^6} r^2.
\end{equation}
These exact charge density and fluid density functions were considered in our study of charged perfect fluid spheres \cite{Dev1}.  The choice of anisotropy has no effect on the metric function  $g_{rr} = e^{2 \lambda(r)}$. It remains the same as in the charged perfect fluid case.  The addition of anisotropy to the system however results in an extra degree of freedom in the equation that determines   $g_{tt} = e^{2 \nu(r)}$. The equation that determines $e^{\nu(r)} \equiv \zeta(r) $ is now written as 

\begin{equation}
\label{zeta2}
\left( \frac{1}{r} e^{-\lambda} \zeta^{\prime} \right)^{\prime}  = \left( \frac{11}{5} - \frac{b}{5}  + j\right) \frac{Q^2}{R^6} r e^{\lambda} \zeta.
\end{equation}
 In the isotropic case we found  solutions for $\zeta(r)$  by varying $b$. The addition of anisotropy allows us to fix  $b$ and vary  $j$ or vice versa when generating  solutions for $\zeta(r)$.
 Thus for each $b$ there are now a very large number of solutions for $\zeta(r)$. Since $b$ appears on equal footing as $j$ in  (\ref{zeta2}) (with the exception of a sign difference) we found the effect of fixing $b$ and varying $j$ to be similar to the results for the isotropic case when $b$ was varied. 
 
\noindent  A summary of our results are as follows:
 \begin{itemize} 
 \item We considered in detail the effects of varying $j$ on a spheres with $b =0$, $b = 1$ and $b = 6$. We found that by varying $j$ for a fixed $b$ a set of critical values curves can be generated that are similar to critical values curves in the isotropic case when we varied $b$. In particular we found that for a sufficiently large positive $j$ we can generate critical values curves with values of $M/R$ greater than those given by the   Andr\'{e}asson           formula.

\item We found in the $b =0, 1$ and $6$ models when $j$ is large and negative the critical values curve becomes quasi-periodic and this behavior gives more than one value for the extremal condition $M/R = Q/R$. 

\item We also studied solutions with the condition  $(2 - \ba + j) = 0$ for both $\ba < 0 $ and $\ba > 0$,  ($\ba = b/5 - 1/5$). Interestingly the critical	values curves in both cases are similar to each other and for a sufficiently large $|\ba|$ we can generate  curves with critical values of $M/R$ greater than those from the  Andr\'{e}asson           formula.

\item Finally we considered solutions with the condition $(2 -\ba +j) = a^2$, and $a^2 = 4 \ba$. Here we were able to reduce the critical values equations to polynomials. This allowed us to  find $M/R$ values for all $0 < Q/R < 1$. When $\ba > 2$ we found that the critical values curves have values of $M/R$ that are larger than those from  the  Andr\'{e}asson           formula.\end{itemize}

\end{document}